\documentclass[prd,aps,preprint,amsmath,nofootinbib,amssymb,eqsecnum,showkeys,tightenlines]{revtex4-1}

\usepackage{amsmath, latexsym, amssymb, hyperref, graphicx, color, multirow, makecell, diagbox}
\usepackage[table]{xcolor}
\usepackage{tabularx}
\usepackage[T1]{fontenc}
\usepackage[capitalize]{cleveref}
\definecolor{nicered}{rgb}{.7,.1,.1}
\definecolor{nicegreen}{rgb}{.1,.5,.1}
\definecolor{darkblue}{rgb}{0,0,.5}
\hypersetup{colorlinks, citecolor=nicegreen,linkcolor=nicered, urlcolor=darkblue}
\usepackage{multirow}
\usepackage{slashed}
\numberwithin{equation}{section}
\usepackage{verbatim}

\begin{document}
\preprint{}

\title{The quirk trajectory}

\author{Jinmian Li$^{1}$}
\email{jmli@scu.edu.cn}

\author{Tianjun Li$^{2,3}$}
\email{tli@mail.itp.ac.cn}

\author{Junle Pei$^{2,3}$}
\email{peijunle@mail.itp.ac.cn}

\author{Wenxing Zhang$^{2,3}$}
\email{zhangwenxing@mail.itp.ac.cn}

\affiliation{$^1$ College of Physics, Sichuan University, Chengdu 610065, China}
\affiliation{$^2$CAS Key Laboratory of Theoretical Physics, Institute of Theoretical Physics, Chinese Academy of Sciences, Beijing 100190, China}
\affiliation{$^3$ School of Physical Sciences, University of Chinese Academy of Sciences,
No.~19A Yuquan Road, Beijing 100049, China}

\begin{abstract}

We for the first time obtain the analytical solution for the quirk equation of motion 
in an approximate way. Based on it, we study several features of quirk trajectory 
in a more precise way, including quirk oscillation amplitude, number of periods, 
as well as the thickness of quirk pair plane. Moreover, we find an exceptional case 
where the quirk crosses at least one of the tracking layers repeatedly. Finally, 
we consider the effects of ionization energy loss and fixed direction of infracolor 
string for a few existing searches. 

\end{abstract}

\maketitle

\section{Introduction}\label{sec:intro}

Solutions to the gauge hierarchy problem of the Standard Model (SM) of particle physics such as supersymmetry and composite Higgs models usually 
predict a colored top partner with mass around TeV scale.
They have been challenged by the null results of LHC searches so far. 
Theories of neutral naturalness~\cite{Curtin:2015bka} aim to address the gauge hierarchy problem 
without introducing colored states, thus relieve the tension with the LHC searches. 
This class of models include folded supersymmetry~\cite{Burdman:2006tz,Burdman:2008ek}, quirky little Higgs~\cite{Cai:2008au}, twin Higgs~\cite{Chacko:2005pe,Craig:2015pha,Serra:2019omd}, minimal neutral naturalness model~\cite{Xu:2018ofw} and so on. In those models, some new $SU(N)$ gauge symmetries are introduced in addition to the SM gauge group. 
A particle,  which is charged under both the SM electroweak gauge group and the new confining $SU(N)$ gauge group
and has mass much larger than the confinement scale ($\Lambda$) of the $SU(N)$, is dubbed as quirk. 
At colliders, the quirk can only be produced in pairs due to the conserved $SU(N)$ symmetry. 
The infracolor force ($F_{s}$, interaction induced by the $SU(N)$ gauge bosons) between two quirks 
will lead to non-conventional signals in the detector. The manifestation of the quirk signal is 
strongly dependent on $\Lambda$ due to $F_s \propto \Lambda^2$~\cite{Kang:2008ea}. 

Throughout the work, we focus on quirk with mass around the EW scale, motivated by the gauge hierarchy problem. 
For $\Lambda \gtrsim \mathcal{O}(10)$ MeV, the strong infracolor force will lead to intensive oscillations in quirk motion. The {quirk-pair} system will lose kinetic energy quickly via photon and hidden glueball radiation. Subsequently, they will annihilate almost promptly into the SM particles after production. Such quirk signals can be searched through resonances in the SM final states~\cite{Cheung:2008ke,Harnik:2008ax,Harnik:2011mv,Fok:2011yc,Chacko:2015fbc,Capdevilla:2019zbx}. 
When $\Lambda \in [10~\text{keV}$, $10~\text{MeV}]$, the quirk pair oscillation amplitude is microscopic ($\lesssim \mu m$), while the energy loss due to photon and hidden glueball radiation is not efficient. The electric neutral quirk-pair system will leave a straight line {inside} the tracker, which will be reconstructed as a single ultra-boosted charged particle with a high ionization energy loss (different from conventional heavy stable charged particle). This signal was looked for at the Tevatron~\cite{Abazov:2010yb}. 
As for $\Lambda \lesssim \mathcal{O}(10)$ eV, the infracolor force is too small to push the quirk out of its helical trajectory, because finite spacial resolution is considered and typical track reconstruction allows the $\chi^2/\text{DOF}$ in fitting as large as 5~\cite{CMS-PAS-EXO-16-036}.  Then the trajectory of each quirk can be reconstructed as normal track {in} detector. The signal can be constrained by conventional heavy stable charged particle searches at the LHC~\cite{CMS-PAS-EXO-16-036,Aaboud:2016uth,Farina:2017cts}.

The scenario with {$\Lambda \in [100~\text{eV}$, $10~\text{keV}]$} is most interesting, since the quirk pair oscillation amplitude can be macroscopic ($\sim$ cm) and the tracker can resolve hits {of} two quirks on each tracking layer. In this case, the quirk trajectory is no longer a helix. The hits caused by the quirks on tracking layers will be completely ignored in conventional event reconstruction at the LHC.  
Meanwhile, as we will show later, the quirk energy deposit in electromagnetic/hadronic calorimeter (ECal/HCal) is usually small within the time period of bunch crossing (25 ns) at the LHC. 
As a result, the quirk simply behaves as missing transverse energy and will be constrained by mono-jet searches at the LHC~\cite{CMS-PAS-EXO-16-037,Aaboud:2016tnv,Farina:2017cts} if the quirk pair is produced recoiling against an energetic initial state radiated (ISR) jet. In fact, there are still many features of quirk trajectory that we could use to help identifying the quirk signal in detector. Ref.~\cite{Knapen:2017kly} innovatively pointed out that if the quirk-pair system is relatively boosted and the infracolor force is much stronger than Lorentz force ($B=4$ T inside the CMS detector), the hits of quirk pair in the detector will almost lie on a plane with deviation less than $\sim \mathcal{O}(100)~\mu m$. They found that the coplanar hits search suffers little background while {maintaining} very high signal efficiency. 
What is more, the quirk carries electric charge and moves slowly when two quirks are widely separated (the typical velocity of {quirk-pair system} is $\sim 0.1$ for quirk pair transverse momentum $p_T>100$ GeV and quirk mass ${m} \sim 100$ GeV). The information of  relatively large ionization energy loss at each hit in the tracker~\cite{ATLAS:2011gea} can be used to further improve the coplanar hits search~\cite{Li:2019wce}.  
On the other hand, if the kinetic energy of the {quirk-pair system} is small, the ECal/HCal of detector will be able to stop the quirk pair. Then, after a long time oscillation inside the calorimeter, the quirk pair {annihilates} eventually. If the annihilation happens at a time when there are no active $pp$ collisions, the signal would be captured by stopped long-lived particles searches at the LHC~\cite{Evans:2018jmd,Aad:2013gva,Sirunyan:2017sbs}. 
 
In this paper we will present an improved understanding of the quirk trajectory inside detector. In particular, we provide an approximate analytical solution for the quirk equation of motion (EoM) for the first time. Based on this solution, we can obtain a more precise expression for the quirk pair oscillation amplitude as well as the number of periods of quirk's motion inside the detector. 
The thickness of quirk pair  plane will be discussed in details. Its dependence on kinetic variables, the confinement scale as well as the quirk charge will be given. We also find that there is great possibility for each quirk traveling though the same tracking layer more than once, leading to multiple resolvable hits on the layer. Note that the coplanar search proposed in Ref.~\cite{Knapen:2017kly} becomes less efficient in this situation. Finally, we briefly {discuss how} the existing searches for the quirk pair with macroscopic oscillation {amplitude} will change if one includes the ionization energy loss in solving quirk EoM or assumes the direction of infracolor string is fixed (The latter assumption seems to be taken in Ref.~\cite{Evans:2018jmd}. ).

In the following, we will consider the infra-gauge group as $SU(3)$. Without specification, the quirk ($\mathcal{Q}$) is fermion with SM quantum numbers ($3,1,2/3$). The colored quirk contradicts the principle of neutral naturalness, but it has large production rate at the LHC and may manifest itself first. The SM quantum numbers only affect the production channel of the quirk, which will lead to different distributions of quirk initial momenta. The analysis proposed in this work is fully applicable to quirks with other quantum numbers.
%Even though most of our discussions do not depend on the specific quirk quantum number and spin, for illustration purpose, we will consider the infra-gauge group as $SU(3)$. Without specification, the quirk ($\mathcal{Q}$) is fermion with SM quantum numbers ($3,1,2/3$).  
%{\color{red} The quirk is chosen to be colored which contradicts the principle of neutral naturalness, but this study is phenomenological in nature and it makes sense to maximize the effects by considering QCD production.}
%The SM quantum numbers mainly affect the production channel of quirk, which {leads} to different distributions of quirk initial momenta. For our choice, the quirk is dominantly produced through QCD interaction. 
Note {that due} to color confinement, only the quirk-quark bound state is observable in experiment and its electric charge can be either $\pm 1$ or zero. It was {found} by Pythia8~\cite{Sjostrand:2007gs} simulation that around 30\%~\cite{Knapen:2017kly} of quirk-quark bound states have charge $\pm 1$. Only the quirk bound states with non-zero electric charge are considered in this work. 
In our Monte Carlo simulation, the events are generated with MG5\_aMC@NLO~\cite{Alwall:2014hca} framework, where the simplified quirk model is written in UFO format by FeynRules~\cite{Alloul:2013bka}, the parton shower and hadronization of the ISR jet {are} implemented by Pythia8. However, the QCD parton shower and hadronization of colored quirk are ignored, since they will not significantly change the kinetic energy of the quirk.  We adopt the CMS detector configuration,  simulated the ionization energy loss of both SM particles and the quirk inside the detector, as well as including the pile-up events. The detailed introduction about the simulation can be found in our previous work~\cite{Li:2019wce}. 

The paper is organized as following. The Section \ref{sec:motion} is devoted to solve the quirk EoM analytically, either with or without including the external forces. The oscillation amplitude, period number as well as the plane thickness of quirk pair trajectory in the tracker are discussed in the Sections \ref{sec:amp}, \ref{sec:period} and \ref{sec:thickness}, respectively. In Section \ref{sec:morehit}, we find an exceptional case where the quirk crosses at least one of the tracking layers repeatedly.
We study the effects of ionization energy loss and fixed $\hat{s}$ in obtaining quirk bounds from existing searches
respectively in the Sections \ref{sec:mono} and \ref{sec:fixs}, and conclude our work in Section \ref{sec:conclude}. 
Moreover, the technical details are given in Appendices \ref{sec:zi}, \ref{sec:dc}, \ref{sec:dvalid}
and \ref{app:hits}.

%%%%%%%%

%%

\section{Quirk motions inside detector}\label{sec:motion}

The quirk equation of motion (EoM) inside detector is given by~\cite{Kang:2008ea}
\begin{align}
\frac{\partial ({m} \gamma \vec{v})}{\partial t}
&=\vec{F}_{s}+\vec{F}_{ext}~,\label{eq::move}\\
\vec{F}_{s}&=-\Lambda^2\sqrt{1-\vec{v}_{\perp}^{2}} \hat{s}-\Lambda^2 \frac{v_{ \|} \vec{v}_{\perp}}{\sqrt{1-\vec{v}_{\perp}^{2}}}~,\label{eq::fs}\\
\vec{F}_{\text{ext}}&=q\vec{v}\times \vec{B}-\langle \frac{dE}{dx}\rangle \hat{v}~,
\end{align}
where {$\gamma=1/\sqrt{1-\vec{v}^2}$,} $v_{ \|}=\vec{v}\cdot\hat{s}\nonumber$ and $\vec{v}_{\perp}=\vec{v}-v_{ \|}\hat{s}\nonumber$ with $\hat{s}$ being a unit vector along the string pointing outward at the endpoints. $\vec{F}_s$ corresponds to the infracolor force and is described by the Nambu-Goto action, where $\Lambda$ is the confinement scale. $\vec{F}_{\text{ext}}$ represents the external forces including Lorentz force and the effects of ionization energy loss for charged quirk propagating in magnetic field and through materials, respectively. Note that we have ignored several sub-dominating energy loss effects such as infracolor glueball and photon radiation, as well as hadronic interaction with detector. 

To solve Eq.~\ref{eq::move}, we have to consider both the quirk pair centre of mass (CoM) frame and the lab frame. In the CoM frame, $\hat{s}$ is approximately parallel to the vector difference between positions of the two quirks (this is only true for $\Lambda^2 \gg F_{\text{ext}}$, see Ref.~\cite{Kang:2008ea}). However, the CoM frame itself is changing all the time due to effects of $\vec{F}_{\text{ext}}$, which is related to quirk velocity in the lab frame. The procedures of numerically solving the EoM by slowly increasing the time with small steps were introduced in Ref.~\cite{Li:2019wce}. In what follows, we will provide a detailed analytical solution for the quirk EoM which is found to be accurate in a wide range of parameter space. 

With $|\vec{F}_{\text{ext}}|\ll\Lambda^2$ (which corresponds to $\Lambda > \mathcal{O}(100)$ eV in the CMS detector), it is reasonable to first ignore $\vec{F}_{\text{ext}}$ if we do not plan to collect the information of ionization energy loss, and take the contribution of $\vec{F}_{\text{ext}}$ as a correction. 

\subsection{Kinematics without $\vec{F}_{\text{ext}}$}
Taking $\vec{F}_{\text{ext}} =0$ in Eq.~\ref{eq::move}, the trajectories of two quirks will lie exactly on the plane (denoted by A ) constructed by $\vec{P}_{o1}$ and $\vec{P}_{o2}$, {with $\vec{P}_{oi}$ ($E_{oi}$) corresponding} to the $i$th quirk initial momentum (energy) in the lab frame. 
Denoting $\hat{e}_x$ and $\hat{e}_y$ as the unit vectors of $\left(\vec{P}_{o1}/{E_{o1}}-\vec{P}_{o2}/{E_{o2}}\right)$ and $\left(\vec{P}_{o1}-\vec{P}_{o1}\cdot\hat{e}_x \hat{e}_x\right)$ respectively,
the $\hat{s}_i~(i=1,2)$ in Eq.~\ref{eq::fs} is the same as either of $\pm \hat{e}_x$ all the time and {reverses} for each time when two quirks cross each other during the oscillation. 
We can use  $\rho_i=\frac{\vec{P}_{oi}\cdot\hat{e}_x}{{m}}$, $h=\frac{\vec{P}_{o1}\cdot\hat{e}_y}{E_{o1}}$ and $l=\frac{\sqrt{1-h^2}\Lambda^2t}{{m}(\rho_1-\rho_2)}$ to define
\begin{align}
\vec{r}_{o1}(l)&=\frac{{m}}{\Lambda^2}\left( \sqrt{1+\rho_1^2}-\sqrt{1+\left(\rho_1-l(\rho_1-\rho_2) \right)^2}\right)\hat{e}_x+
\frac{{m}}{\Lambda^2}\frac{h(\rho_1-\rho_2)l}{\sqrt{1-h^2}}\hat{e}_y~,\\
\vec{r}_{o2}(l)&=\frac{{m}}{\Lambda^2}\left(\sqrt{1+\left(\rho_2-l(\rho_2-\rho_1) \right)^2}- \sqrt{1+\rho_2^2}\right)\hat{e}_x+
\frac{{m}}{\Lambda^2}\frac{h(\rho_1-\rho_2)l}{\sqrt{1-h^2}}\hat{e}_y~.
\end{align}

According to the Eq.~\ref{eq::move}, the trajectories of two quirks in the lab frame will be
\begin{align}
\vec{r}_{1}(l)&=[l]\vec{r}_{o1}(1)+\frac{1+(-1)^{[l]}}{2}\vec{r}_{o1}(\bar{l})+\frac{1-(-1)^{[l]}}{2}\vec{r}_{o2}(\bar{l})~,\label{eq::r1}\\
\vec{r}_{2}(l)&=[l]\vec{r}_{o2}(1)+\frac{1-(-1)^{[l]}}{2}\vec{r}_{o1}(\bar{l})+\frac{1+(-1)^{[l]}}{2}\vec{r}_{o2}(\bar{l})~,\label{eq::r2}
\end{align}
where $[l]$ is {floor} of $l$ and $\bar{l} \equiv l-[l]$. {It is straight forward to find that $\vec{r}_{1}(l)=\vec{r}_{2}(l)$ when $l$ is an integer and the time interval between $l$ and $l+1$ is
\begin{align}
{t_p=0.0658\frac{(\rho_1-\rho_2)}{\sqrt{1-h^2}}\frac{{m}}{\text{[100~GeV]}}\frac{\text{[keV]}^2}{\Lambda^2}~\text{[ns]}}~.\label{eq::dt}
\end{align}
}

\subsection{Kinematics with uniform magnetic field}\label{sec:kinB}

With an uniform magnetic field $\vec{B}$ in the lab frame, the Lorentz forces on two quirks will push them out of the plane A. 
Then the motions of two quirks can be decomposed into two parts: one is parallel to plane A and the other is perpendicular to plane A. In the limit of $F_{\text{Lorentz}} \ll\Lambda^2$, the motions parallel to plane A can be approximated by Eqs.~\ref{eq::r1} and \ref{eq::r2}. 
By adding the motion perpendicular to plane A, the trajectories of two quirks in the lab frame can be expressed as
\begin{align}
\vec{r}_{i}^{~\prime}(l)&=\vec{r}_{i}(l)+z_i(l)\hat{e}_z~,
\end{align}
where $\hat{e}_z=\hat{e}_x\times \hat{e}_y$ with $\hat{e}_x$ and $\hat{e}_y$ defined as before. 

It will be easier to calculate $z_i$ in the {approximately invariant} CoM frame. Each of the initial quirks momenta ($\vec{P}_{oi}^\prime$) in the CoM frame is given by $|\vec{P}_{oi}^\prime|={m}\rho$ with
\begin{align}
&\rho=\sqrt{\frac{(E_{o1}+E_{o2})^2-(\vec{P}_{o1}+\vec{P}_{o2})^2}{4{m}^2}-1}~.\label{ap::rhoc}
\end{align}  

The quirks motions described by Eqs.~\ref{eq::r1} and \ref{eq::r2} in the lab frame can be boosted into the CoM frame and expressed as
\begin{align}
\vec{v}_{c1}(g)&=-\vec{v}_{c2}(g)=(-1)^{[g]}
v(\bar{g})\hat{e}_{xc}~,\label{ap::vc}\\
\vec{r}_{c1}(g)&=-\vec{r}_{c2}(g)=(-1)^{[g]} r(\bar{g})\hat{e}_{xc}~,\label{ap::rc}\\
v(g)&= \frac{\rho(1-2g)}{\sqrt{1+\rho^2(1-2g)^2}}~, \\
r(g)&=\frac{{m}}{\Lambda^2}\left(\sqrt{1+\rho^2}-\sqrt{1+\rho^2(1-2g)^2} \right)~, \label{ap::rcc}
\end{align}  
where $\hat{e}_{xc}$ is the unit vector of $\vec{P}_{o1}^\prime$, $\hat{e}_{yc}=\hat{e}_{z}\times\hat{e}_{xc}$, $g=\frac{\Lambda^2 t_c}{2{m}\rho}$, $\bar{g}=g-[g]$, and $t_c$ is time in the CoM frame. Note that we have used subscript $c$ for variables in the CoM frame.

Since the velocities of both quirks are approximately along $\pm\hat{e}_{xc}$ in the CoM frame, the {$\hat{e}_{z}$-component} of $\vec{F}_{\text{ext}}$ is {induced} by the $\hat{e}_{yc}$-component of the  magnetic {field}  and the $\hat{e}_{z}$-component of the electric field, which are
\begin{align}
B_c&=\frac{\left( {1-\beta_o^2}\right) \vec{P}_{o1}\cdot\hat{e}_{x\prime}\vec{B}\cdot\hat{e}_{y\prime}-{\left( \vec{P}_{o1}\cdot\hat{e}_{y\prime}-\beta_o E_{o1}\right)\vec{B}\cdot\hat{e}_{x\prime}} }{{m}\rho \left( {1-\beta_o^2}\right)}~,\\
E_z&=-\frac{\beta_o \vec{B}\cdot\hat{e}_{x\prime} }{\sqrt{1-\beta_o^2}}~,
\end{align}
where $\vec{\beta}_o=\frac{\vec{P}_{o1}+\vec{P}_{o2}}{E_{o1}+E_{o2}}$, $\beta_o=|\vec{\beta_o}|$, $\hat{e}_{y\prime}$ and $\hat{e}_{x\prime}$ are the unit vectors of $\vec{\beta}_o$ and $( \vec{P}_{o1}-\vec{P}_{o1}\cdot\hat{e}_{y\prime}\hat{e}_{y\prime})$, respectively.

Defining 
\begin{align}
z^{+}(g)&\equiv \frac{1}{2}\left(z_1(g)+z_2(g) \right)~, \\
z^{-}(g)&\equiv \frac{1}{2}\left(z_1(g)-z_2(g) \right)~,
\end{align}
because electric charges of two quirks have opposite sign and their velocities satisfy Eq.~\ref{ap::vc}, we can infer that $z^{+}(g)$ arises from the Lorentz force caused by $B_c$, and $z^{-}(g)$ arises from the electric field force caused by $E_z$.
The detailed derivations for $z^{+}(g)$ and $z^{-}(g)$ are given in Appendix~\ref{sec:zi}. 
{Considering
$t_{ci}=\frac{t_i-\vec{\beta}_o \cdot \vec{r}_i(l_i)}{\sqrt{1-\beta_o^2}}$,
we have}
\begin{align}
g_i(l)&=\frac{1}{\sqrt{1-\beta_o^2}}\left( \frac{(\rho_1-\rho_2)}{2\rho\sqrt{1-h^2}}l-\frac{\Lambda^2\vec{r}_i(l)\cdot\vec{\beta}_o}{2{m}\rho}\right)~. 
\end{align}
The trajectories of two quirks in the lab frame can be expressed as
\begin{align}
\vec{r}_{1}^{~\prime}(l)&=\vec{r}_{1}(l)+\left(z^{+}(g_1(l))+z^{-}(g_1(l)) \right)\hat{e}_z~,\label{eq::motion1}\\
\vec{r}_{2}^{~\prime}(l)&=\vec{r}_{2}(l)+\left(z^{+}(g_2(l))-z^{-}(g_2(l)) \right)\hat{e}_z~.\label{eq::motion2}
\end{align}

\subsection{Variable substitution and benchmark point} \label{sec:vsub}
We have derived the quirk trajectories in the lab frame in terms of $\vec{B}$, ${m}$, $\Lambda$, $\vec{P}_{o1}$ and $\vec{P}_{o2}$. 
In the following discussion, we will take the configuration of the CMS detector whenever discussing the experimental measurements.  This corresponds to $\vec{B}=4$ T along the $\hat{z}$-axis. The $\vec{P}_{o1}$ and $\vec{P}_{o2}$ parameters will be replaced by  $k_i=|\vec{P}_{oi}/m|~(i=1,2)$, $\alpha$, $\theta$ and $\phi$. The $\alpha$ is the angle between 
$\vec{P}_{o1}$ and $\vec{P}_{o2}$. The $\theta$ stands for the angle between $\vec{P}_{o}=\vec{P}_{o1}+\vec{P}_{o2}$
and $\hat{z}$. The $\phi$ is related to the angle between $\vec{P}_{o1} \times \vec{P}_{o2}$ and  $\vec{P}_{o} \times \hat{z}$. To be specific, we have
\begin{align}
\theta&=\begin{cases}
\cos^{-1}(\delta),~~~~~~~~~~~~\delta\ge 0\\
\pi-\cos^{-1}(\delta),~~~~~~~\delta<0
\end{cases}~,\\
\phi&=\begin{cases}
\pi/2-\cos^{-1}(\psi),~~~~\delta\ge 0\\
\cos^{-1}(\psi)-\pi/2,~~~~\delta<0
\end{cases}~,
\end{align}
where $\delta=\frac{\vec{P}_o\cdot \hat{z}}{|\vec{P}_o|}$ and $\psi=\frac{\left( \vec{P}_{o1}\times\vec{P}_{o2}\right) \cdot \left(\vec{P}_{o}\times\hat{z} \right)}{|\vec{P}_{o1}\times\vec{P}_{o2}||\vec{P}_{o}\times\hat{z}|}$.

\begin{figure}[thb]
\begin{center}
\includegraphics[width=0.45\textwidth]{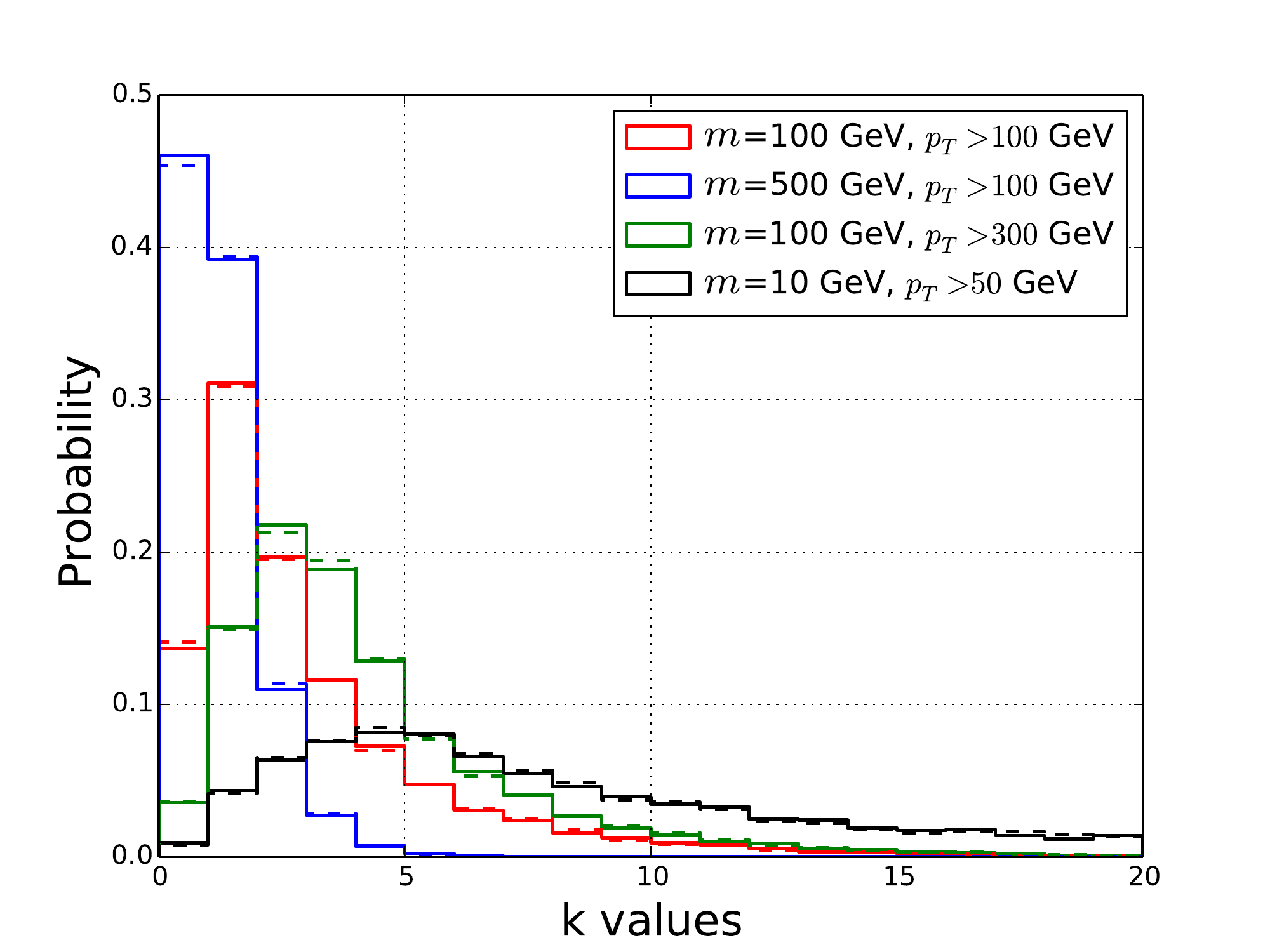}
\includegraphics[width=0.45\textwidth]{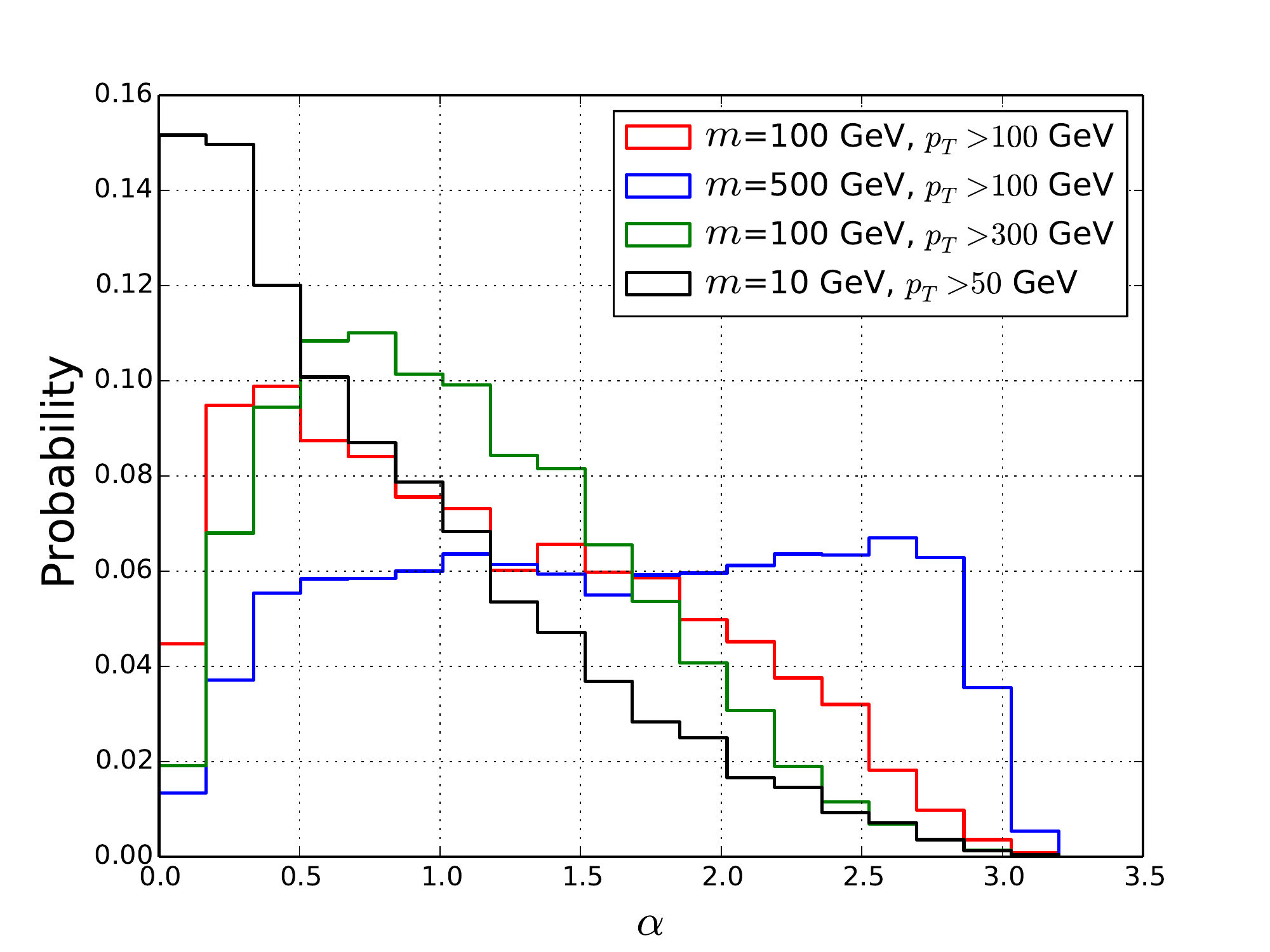}\\
\includegraphics[width=0.45\textwidth]{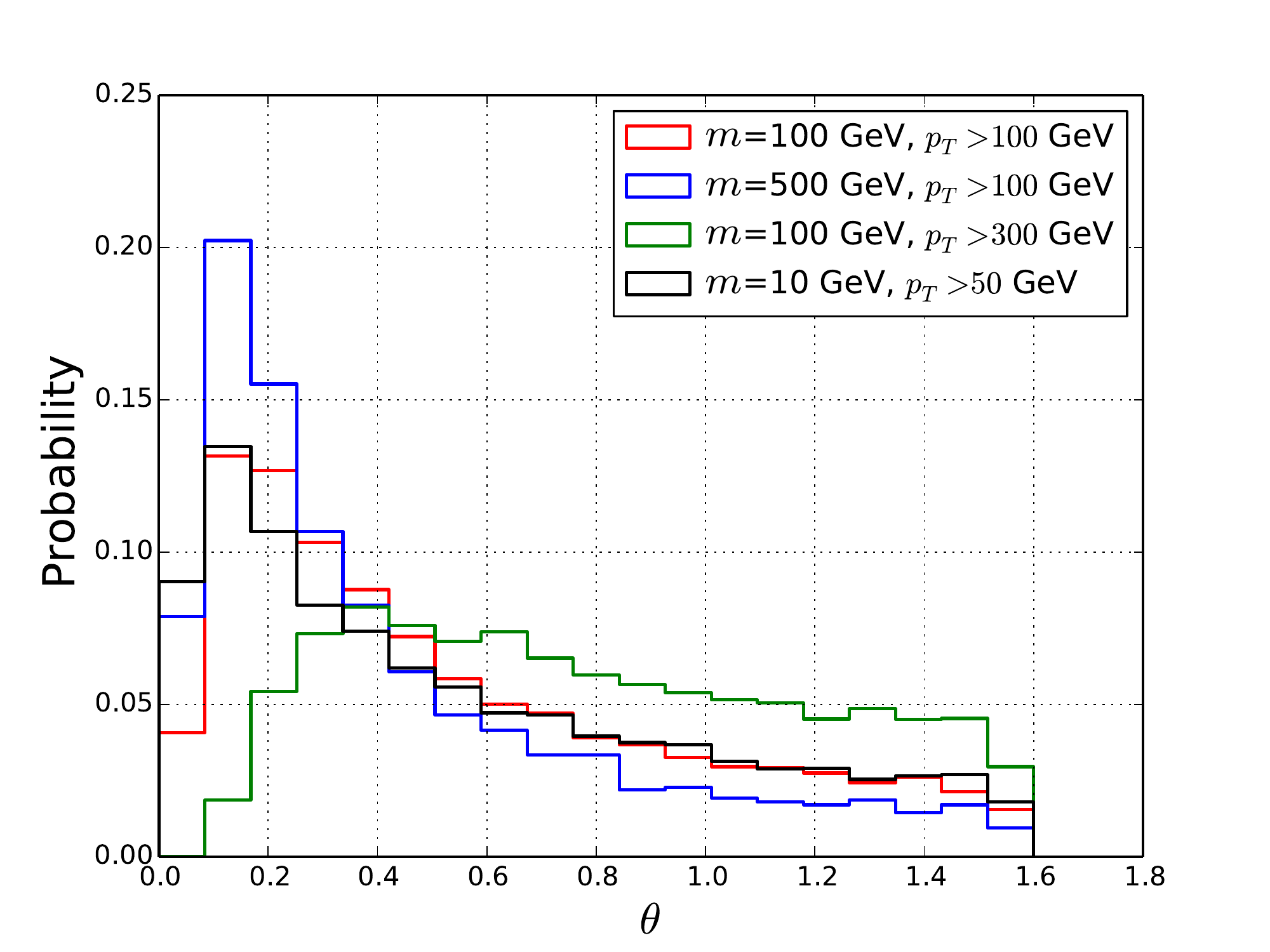}
\includegraphics[width=0.45\textwidth]{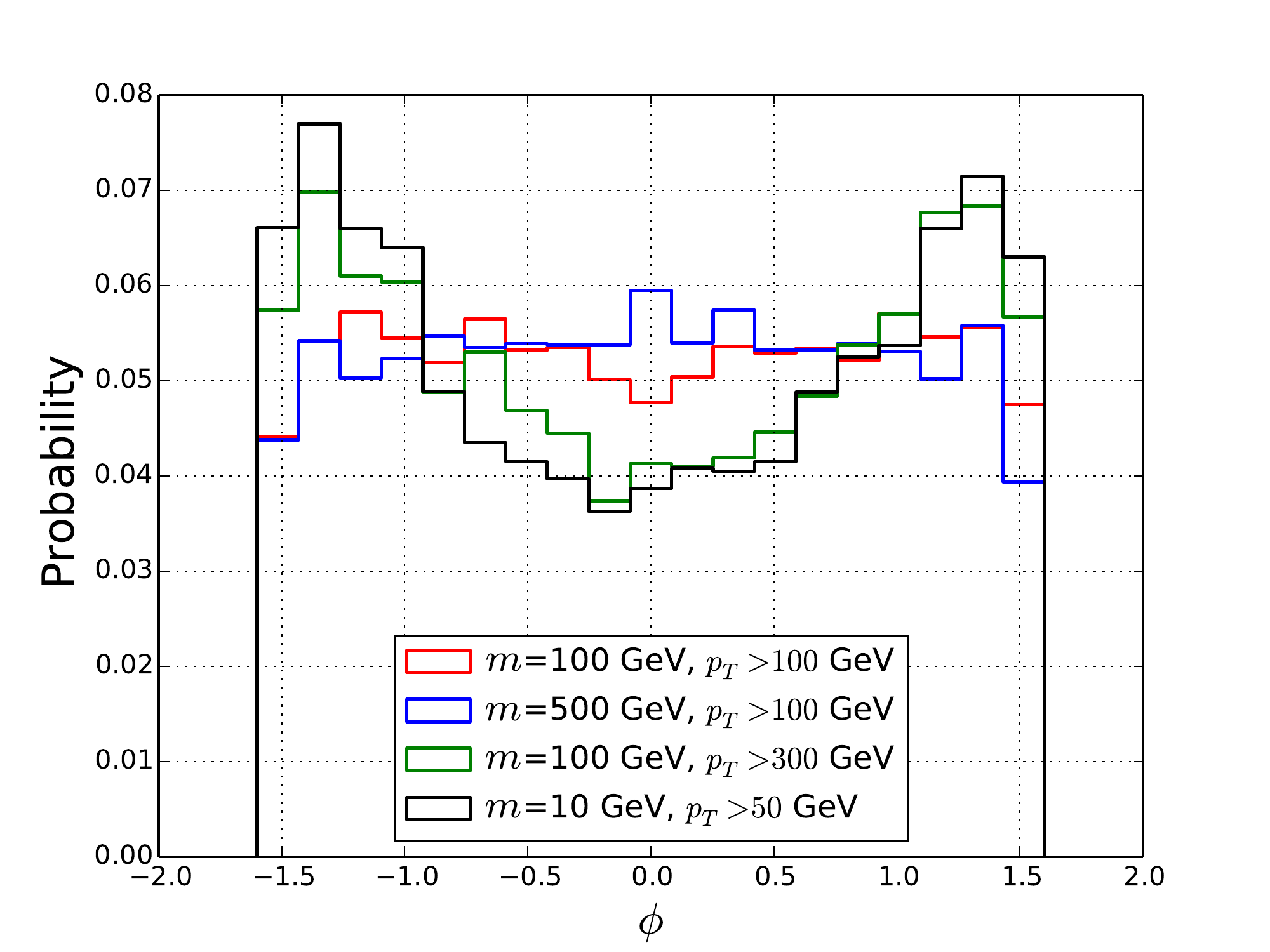}
\end{center}
\caption{Distributions of quirk initial kinematic variables for different quirk masses and quirk-pair system transverse momenta. In the upper-left panel, the solid and dashed lines correspond to $k_1$ and $k_2$, respectively.  }
\label{fig::kins}
\end{figure}

Fig.~\ref{fig::kins} shows the distributions of $k_{1,2}$, $\alpha$, $\theta$ and $\phi$ with different quirk masses and quirk-pair system transverse momenta in our benchmark model. Note that the quirk-pair system is required to be relatively boosted along the transverse direction, in order to produce detectable signals inside the tracker. This is implemented by requiring a hard initial state radiated (ISR) jet that is recoiling against the quirk pair. {Thus, the events of quirk production in the detector can be triggered by either the relatively large missing transverse energy or a hard jet. } Meanwhile, the behaviors of quirk inside the tracker will be also affected by the hardness of the ISR jet. We can find that larger $p_T/{m}$ leads to greater $k_{1,2}$, $\theta$ and $|\phi|$, while rendering smaller $\alpha$.

Since many observables of interests contain complex coefficients, it will be more intuitive to show the results on a benchmark point, which we choose to be 
\begin{align}
k_1=k_2=2.5,~\alpha=\pi/4, ~\theta=0.3, ~ \phi=1.0, ~{\Lambda_0^2} = 4\times 10^5~\text{eV}^2, ~m=100~\text{GeV}.
\end{align}
From Fig.~\ref{fig::kins}, we can see that this point has high probability in most cases. 

\section{Oscillation amplitude in the lab frame} \label{sec:amp}

It is known that the quirks are traveling oscillatingly, with the characteristic amplitude of oscillation in the CoM frame~\cite{Kang:2008ea} $\ell_{c} \sim 2~\text{cm}(\gamma-1) (\frac{{m}}{\text{100~GeV}}) (\frac{\Lambda}{\text{keV}})^{-2}$, {which can be expressed via the new parameters as 
$\ell_{c} \sim 2~\text{cm}(\sqrt{1+\rho^2} -1) \frac{m}{\text{[100~GeV]}}\frac{\text{[keV]}^2}{\Lambda^2}  $
with $\rho  =\sqrt{\frac{\sqrt{1+k_1^2} \sqrt{1+k_2^2} - k_1 k_2 \cos(\alpha) -1}{2}}$ from Eq.~\ref{ap::rhoc}. }The definitions of $k_1$, $k_2$ and $\alpha$ are given in Sec.~\ref{sec:vsub}. In fact, there is another more useful parameter relevant to the width of the quirk oscillation in the lab frame
\begin{align}
L=2 \frac{\mathcal{R}}{\rho}\ell_{c}~,\label{eq::LL}
\end{align}
with $\mathcal{R}=\frac{k_1 k_2 \sin(\alpha)}{\sqrt{k_1^2 +k_2^2 + 2 k_1 k_2 \cos(\alpha)}}$.
{$L$ corresponds to twice the length of the projection of $\ell_{c}$ onto the plane perpendicular to $\vec{\beta}_o$.} It can be calculated immediately that $L=3.8$ cm for our benchmark point.

\section{Number of periods in the tracker} \label{sec:period}

Since two quirks are produced at the same interaction point and traveling oscillatingly, we can define each of the second time when two quirks meet as one period of quirk motion, {corresponding to the $l$ increased by one in our previous discussion.} {The overall shape of the tracker system of the CMS detector is cylindrical and described by $R_{max}=118.5$ cm (3.95 ns in natural units) and $|Z_{max}|=293.5$ cm (9.78 ns)}~\cite{Chatrchyan:2009hg}. So we can calculate the number of periods ($n_T$) that the quirk pair is going through inside the tracker system in 25 ns (bunch crossing time at the LHC)
	\begin{align}
	n_T&=\begin{cases}
	\frac{\text{Min}[25,~9.78/\beta_z]~\text{ns}}{t_p},~~~~~~~\beta_z/\beta_T\ge 2.4768\\
	\frac{\text{Min}[25,~3.95/\beta_T]~\text{ns}}{t_p},~~~~~~~\beta_z/\beta_T<2.4768
	\end{cases}~,
	\end{align}
	where {2.4768=$|Z_{max}|/R_{max}$}, $\beta_z=|\vec{\beta}_o\cdot \hat{z}|$, $\beta_T=\sqrt{\beta_o^2-\beta_z^2}$ and $t_p$ is given in Eq.~\ref{eq::dt}. So we get $n_T \propto \Lambda^2/m$.

\begin{figure}[thb]
\begin{center}
\includegraphics[width=0.3\textwidth]{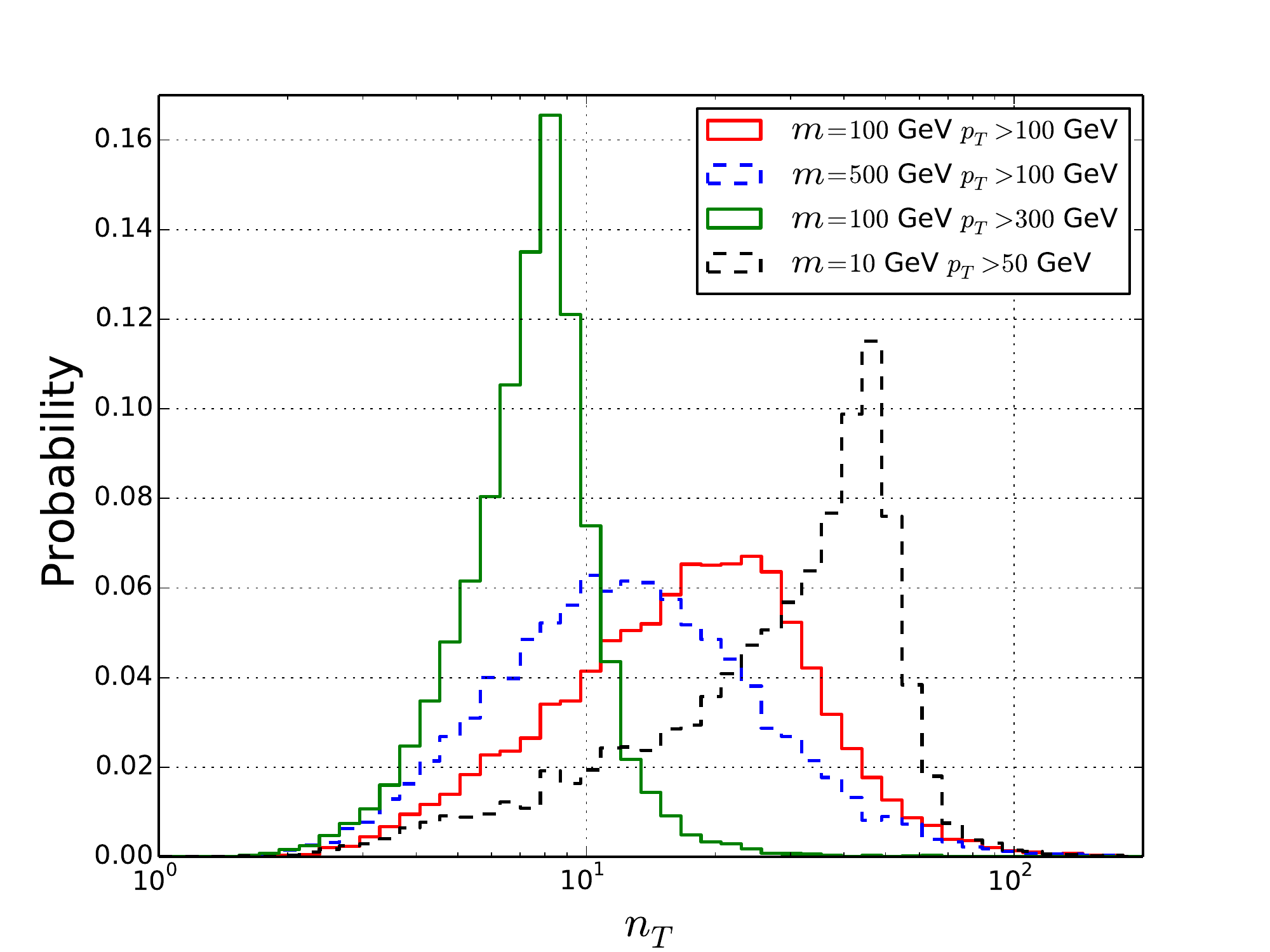}
\includegraphics[width=0.3\textwidth]{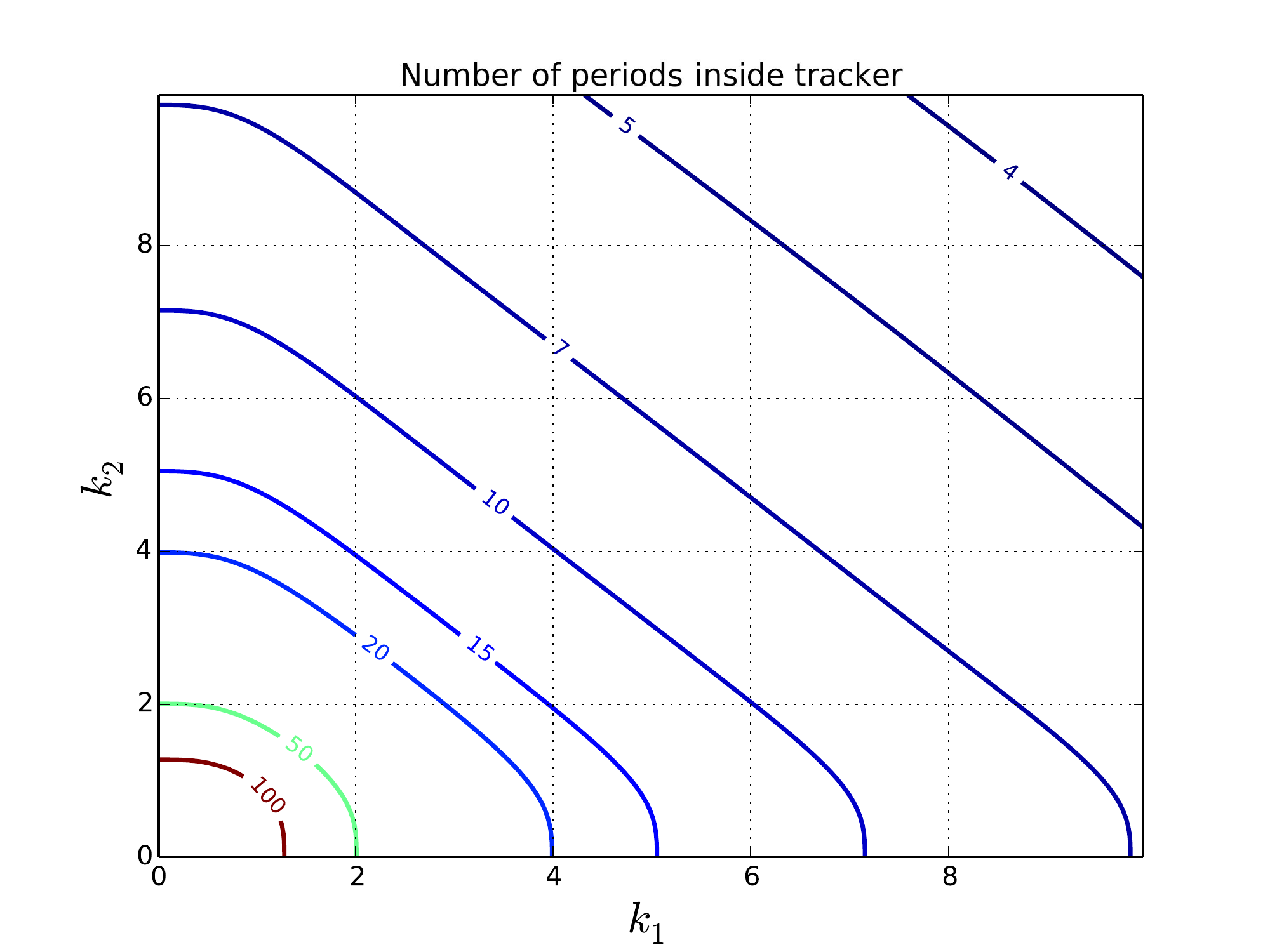}
\includegraphics[width=0.3\textwidth]{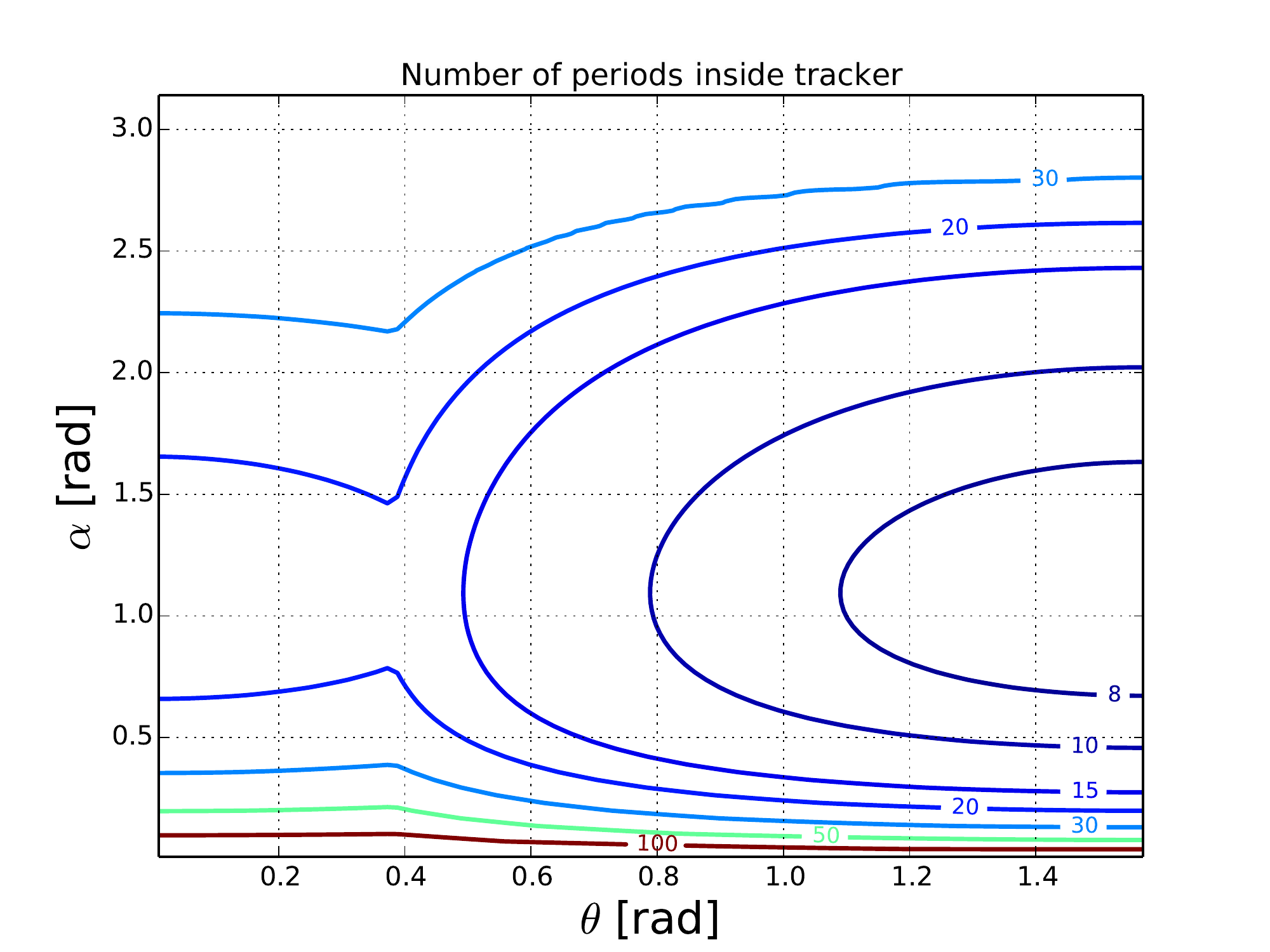}
\end{center}
\caption{\label{fig:kins}  
Left: distributions of number of periods for different quirk masses and quirk-pair transverse momenta (from numerical simulation). Middle and right: the projected (parameters are set at benchmark points if they are not varying) relation between $n_T$ and $k_1-k_2$, and {$\theta-\alpha$} (from analytical calculation). {Here, we have fixed $\Lambda=\Lambda_0$ for all cases}. }
\end{figure}

We can find that our benchmark point is going through 19.5 periods inside the tracker. The $n_T$ is {insensitive to} the $\phi$ parameter.  In Fig.~\ref{fig:kins}, distributions of $n_T$ for different quirk masses and quirk-pair transverse momenta and the dependence of $n_T$ on the $k_1-k_2$ and $\theta-\alpha$ are shown. In the $k_1-k_2$ plane, increasing $k_1$ and $k_2$ lead to larger $t_p$ and shorter time of the quirk pair staying in tracker, and thus smaller $n_T$. In the $\theta-\alpha$ plane, 
 $n_T$ decreases with the increasing small $\alpha$ because of the increased $t_p$. When $\alpha$ is not small, increasing $\alpha$ leads to longer time of the quirk pair staying in tracker and thus larger $n_T$. The quirk pair leaves the tracker by crossing the outermost endcap (barrel) when $\theta<(>) 0.384$ {($\tan[0.384]$=$R_{max}/|Z_{max}|$ )}. Similarly, different quirk masses and quirk-pair transverse momenta lead to different distributions of $n_T$ due to the differences in the parameter space of $k_1-k_2$ and $\theta-\alpha$.

\section{Thickness of quirk pair plane}\label{sec:thickness}

Quirk-antiquirk pair propagating through the tracker system of a detector will leave N hits located at $\vec{h}_i~(i=1,2,\dots,N)$. {It was pointed out firstly in the Ref.~\cite{Knapen:2017kly} that in a wide range of parameter space with $\Lambda \sim \mathcal{O}(1)~\text{keV}$, these hits will largely lie on a plane. The averaged distance of hits to a virtual plane which contains the interaction point (at origin) is calculated by 
\begin{align}
d(\vec{n})=\sqrt{\frac{1}{N-1}\sum_{i=1}^{N}\left( \vec{n}\cdot\vec{h}_i\right) ^2}~,
\end{align}
where  $\vec{n}$ is the normal vector of the plane.
The plane giving the smallest $d(\vec{n})=d_{min}$ is called the quirk pair plane and the $d_{min}$ is called thickness of the quirk pair plane.}

With an uniform magnetic field $\vec{B}$ in the detector, trajectories of the quirk pair can be described 
by Eqs.~\ref{eq::motion1} and \ref{eq::motion2}. {The number of periods that the quirk pair is going through inside the tracker system in 25 ns is $n_T\approx l_0$, where $l_0$ is the integer closest to $n_T$.}

According to the discussions in Sec.~\ref{sec:kinB} and Appendix~\ref{sec:zi}, the quirk pair plane in the CoM frame can be approximately obtained by rotating the $\hat{e}_{xc}-\hat{e}_{yc}$ plane by an angle $(-j\eta)$  around $\hat{e}_{yc}$ and then moving it along $\hat{e}_{z}$ by a distance $c_z$. {Then distances of two quirk trajectories to the plane obtained above are} 
\begin{align}
d_{c1}(g)=&z_1(g)-c_z-(-1)^{[g]}j\eta r(\bar{g})~,\\
d_{c2}(g)=&z_2(g)-c_z+(-1)^{[g]}j\eta r(\bar{g})~,
\end{align}
respectively, so that thickness of quirk pair plane can be approximately expressed as 
\begin{align}
d_c^2&=\frac{1}{2l_0}\int_{0}^{l_0}\left(d_{c1}^2(g)+d_{c2}^2(g) \right)dg\nonumber \\
&\approx \frac{1}{l_0}\sum_{i=1}^{l_0}\left(j-[\frac{i}{2}] \right)^2\int_{0}^{1} \eta^2r^2(g)dg +\int_{0}^{1}
\left(z^{+}_o(g)-c_z \right)^2 dg+\int_{0}^{1}
z^{-}_o(g)^2 dg~. \label{eq:dc2}
\end{align}

Finally, thickness of quirk pair plane in the lab frame can be estimated as (The detailed discussions are provided in Appendix~\ref{sec:dc}.) 
\begin{align}
d&= \sqrt{ \frac{\eta^2r^2(0.5)}{3l_0}\sum_{i=1}^{l_0}\left( j(l_0)-[\frac{i}{2}]\right) ^2+\frac{4}{3}\left( \frac{2{z_o^{+ }}^2(1)}{15}+\frac{{z_{o}^{-}}^2(1)}{5}\right)}  \label{eq::ddsim}\\
 & \approx\sqrt{\left( \frac{1}{3l_0}\sum_{i=1}^{l_0}\left( j(l_0)-[\frac{i}{2}]\right) ^2+\frac{4}{15}\right)d_E^2 +\frac{8}{45} d_B^2}~,\label{eq::dsim}
\end{align}
where we have used the relations of Eq.~\ref{ap::magnitude} in the second line, {and
\begin{align}
j(l_0)&=\frac{1}{l_0}\sum_{i=1}^{l_0}[\frac{i}{2}]~,\\
d_E&={z_{o\prime}^{-}}(1)=\frac{2mqE_z}{\Lambda^4}\rho \sinh^{-1}[\rho]~,\label{ap::de} \\
d_B&={z_o^{+ }}(1)=\frac{2mqB_c}{\Lambda^4}\left(\rho-\tan^{-1}[\rho] \right) ~.\label{ap::db}
\end{align}
}

We provide the validation of Eq.~\ref{eq::dsim} in Appendix~\ref{sec:dvalid}, where we can conclude that our analytic formula for the quirk pair plane thickness matches the numerical result within an order magnitude when quirk pair goes through $\sim \mathcal{O}(1-100)$ periods inside tracker. {The main difference between the quirk pair plane thickness calculated from Eq.~\ref{eq::dsim} ($D_A$) and that obtained from numerical simulation $(D_N)$ in Appendix~\ref{sec:dvalid} attributes to the following reasons. 
First of all, the former uses the shapes of overall quirk trajectories, or positions of infinite points from every parts of quirk trajectories. The latter only uses $\mathcal{O}$(10) positions of hits caused by quirk crossing detector layers in tracker. Besides, we employ the relations of Eq.~\ref{ap::magnitude} to obtain Eq.~\ref{eq::dsim} as an approximation of  Eq.~\ref{eq::ddsim}.}

\begin{figure}[thb]
\begin{center}
\includegraphics[width=0.3\textwidth]{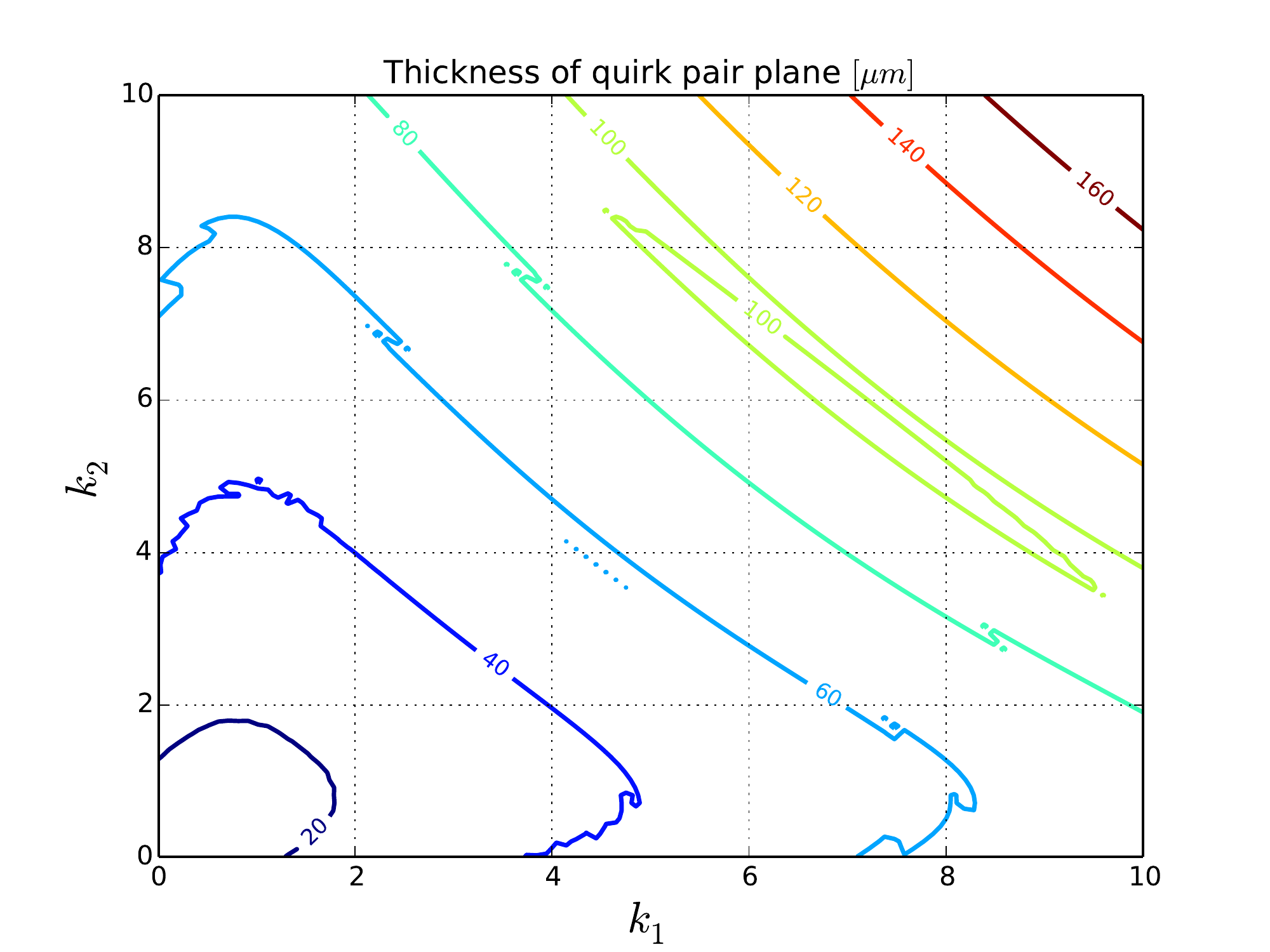}
\includegraphics[width=0.3\textwidth]{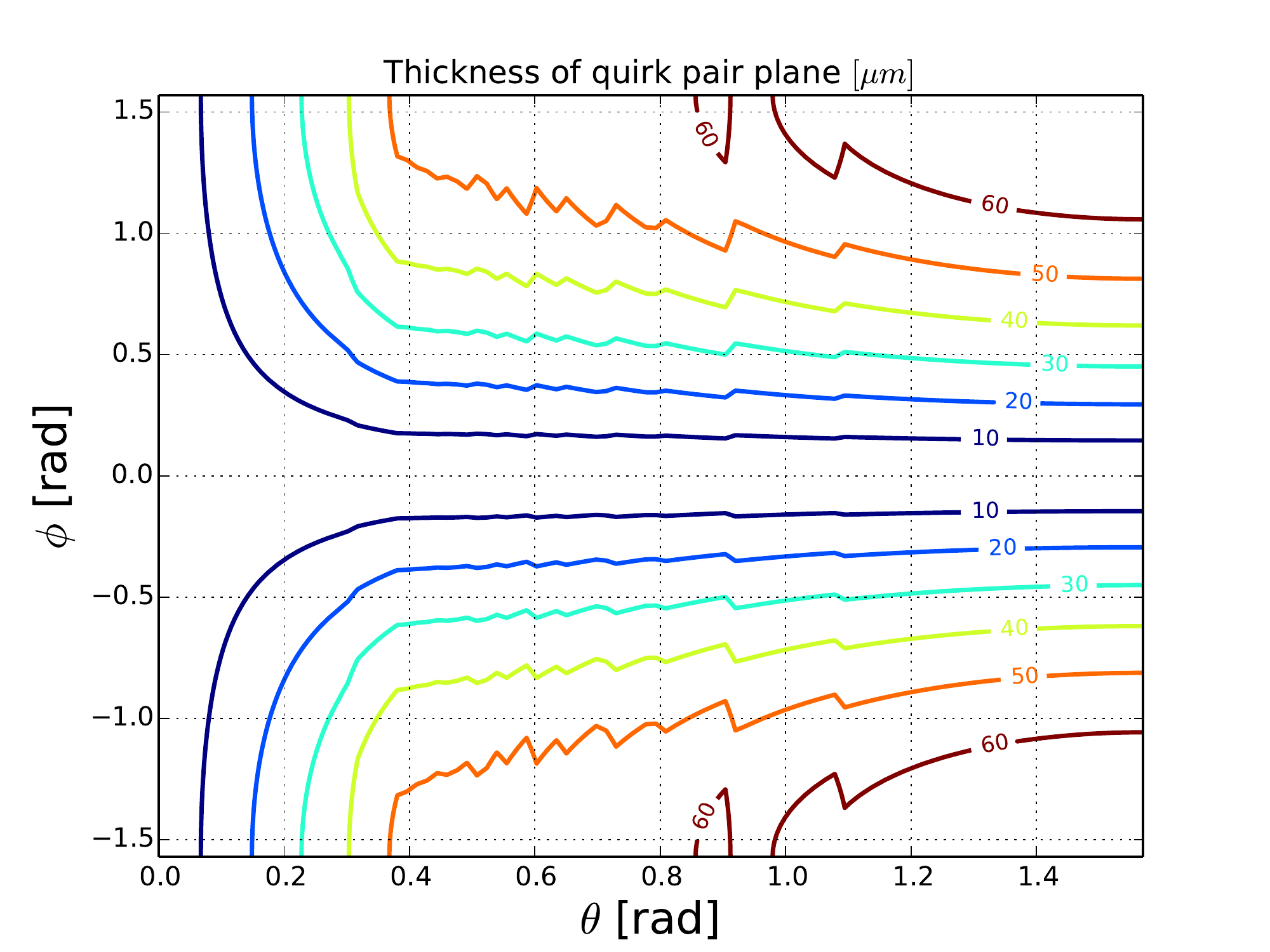}
\includegraphics[width=0.3\textwidth]{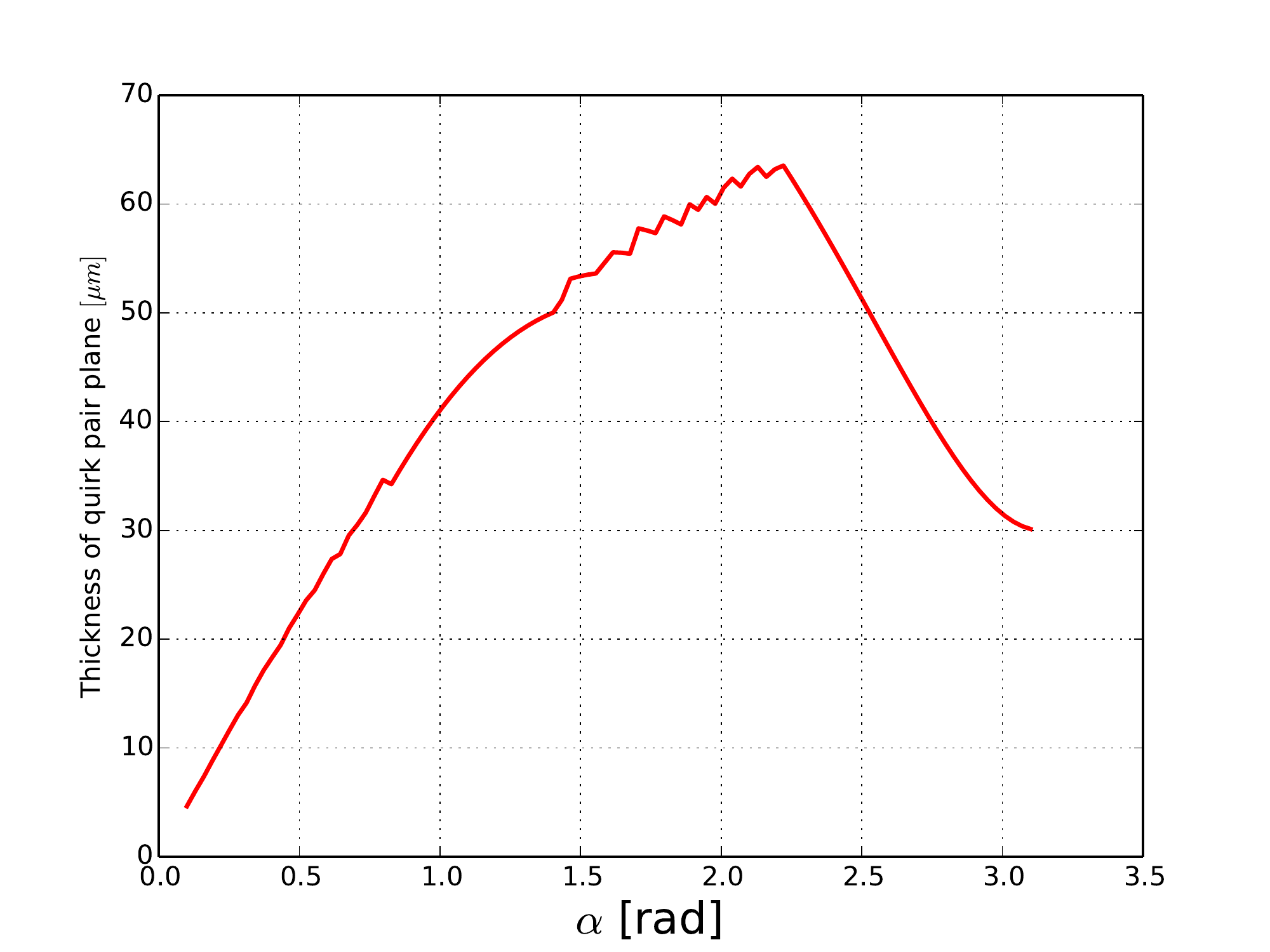}
\end{center}
\caption{Thickness of the quirk pair plane in the projected parameter space: $k_1-k_2$ (left), $\theta-\phi$ (middle), $\alpha$ (right). Parameters are set at benchmark points if they are not considered as varying. {The confinement scale $\Lambda$ is fixed to be $\Lambda_0$ for all cases}. \label{fig::thickness} }
\end{figure}

In Fig.~\ref{fig::thickness}, we show the projected thickness of the quirk pair plane on the parameter space of $k_1-k_2$, $\theta-\phi$ and $\alpha$, respectively, by using the Eq.~\ref{eq::dsim}. The {irrelevant} parameters are set at benchmark point in the projection. For $l_0\sim \mathcal{O}(10)$, $d_E$ plays a more important role in $d$ than $d_B$, {since} $d\approx \sqrt{\mathcal{O}(1) d_E^2+\frac{8}{45}d_B^2}$. In the $k_1-k_2$ plane, increasing $k_1$ and $k_2$ will lead to increased $E_z$ and $\rho$ thus larger $d_E$. Similarly, in the $\theta-\phi$ plane, larger $\theta$ and $|\phi|$ give greater $E_z$ thus larger $d_E$.  The dependence on the $\alpha$ is more complicated. Increasing $\alpha$ will lead to larger $\rho$ parameter while smaller $E_z$. In the small $\alpha$ region, the $\rho$ parameter is dominating. So the quirk pair plane thickness is increased with  $\alpha$. On the other hand, the $E_z$ becomes dominant in the large $\alpha$ region, which gives decreased plane thickness for increasing $\alpha$. 
{Note that the non-smooth behavior of the contours are originated from the fact that $l_0$ is not a smooth function of $t_0$.}

It will be more useful to predict a model or parameter space which has large thickness of quirk pair plane by using Eq.~\ref{eq::dsim} and features in Fig.~\ref{fig::thickness} instead of conducting time consuming numerical simulation. 
In the left panel of Fig.~\ref{fig::dnc}, we plot the distributions of quirk pair plane thickness with varying quirk mass (${m}$), transverse momentum of quirk pair ($p_T$) and confinement scale ($\Lambda$). 
The dependence on $\Lambda$ is obvious: the thickness decreases with increasing $\Lambda$. However, for the considered quirk production process (dominated by $gg\to \mathcal{Q} \mathcal{Q}$), greater $p_T/{m}$ gives {larger $k_{1,2}$, $\theta$ and $|\phi|$, but smaller $\alpha$ and $l_0$.} Thus we can find the thickness dependence on ${m}$ and $p_T$ is mild (we have checked with several parameter {choices which} are not shown in the plot.).  On the other hand, if the quirk pair is produced from a heavy resonant decay $p p \to j Z'(\to \mathcal{Q} \mathcal{Q})$, the quirk pair plane thickness will be much larger for heavier $Z'$. 

\subsection{Charge dependence of the quirk pair plane thickness}
\begin{figure}[t]
\begin{center}
\includegraphics[width=0.45\textwidth]{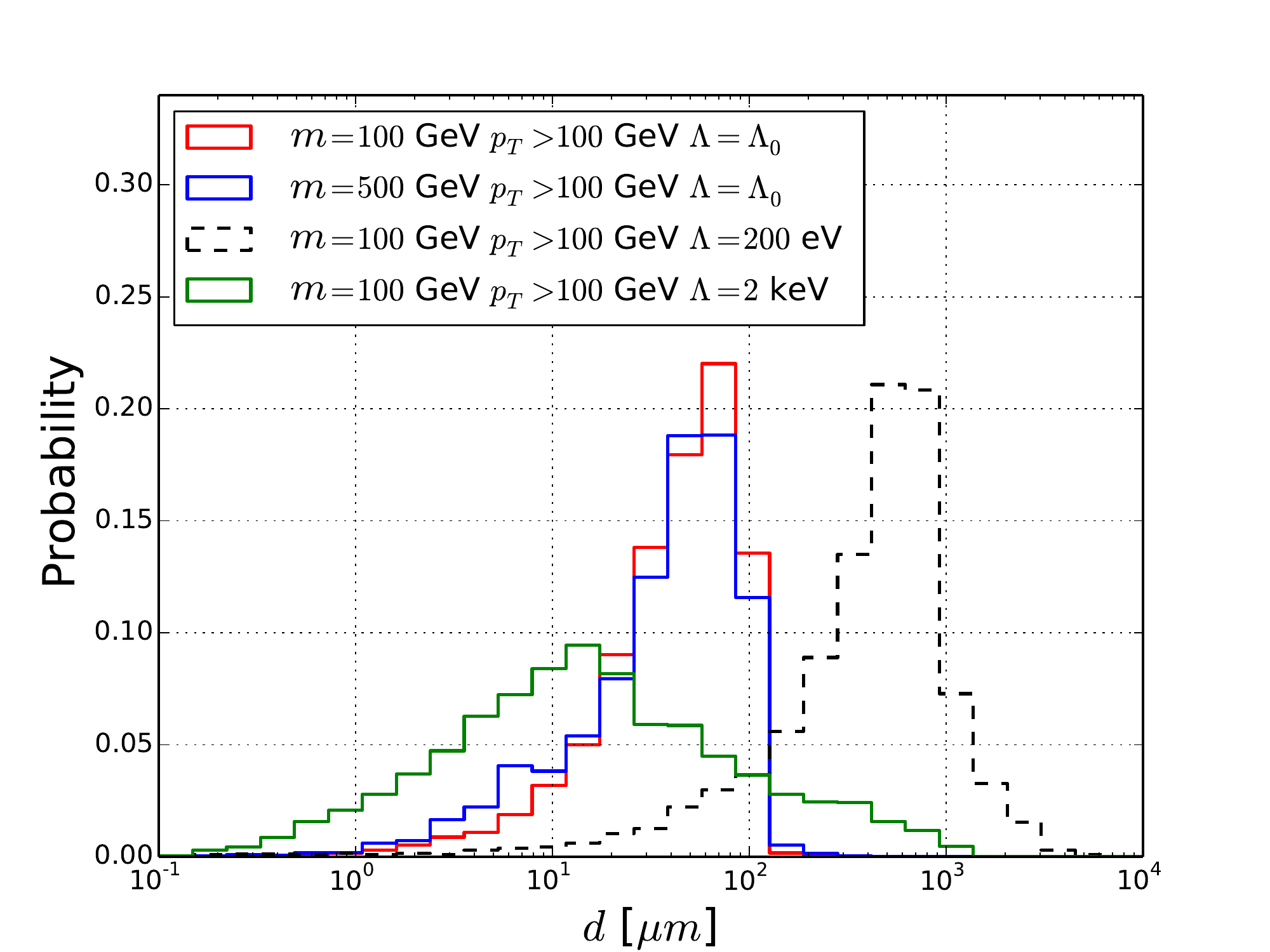}
\includegraphics[width=0.45\textwidth]{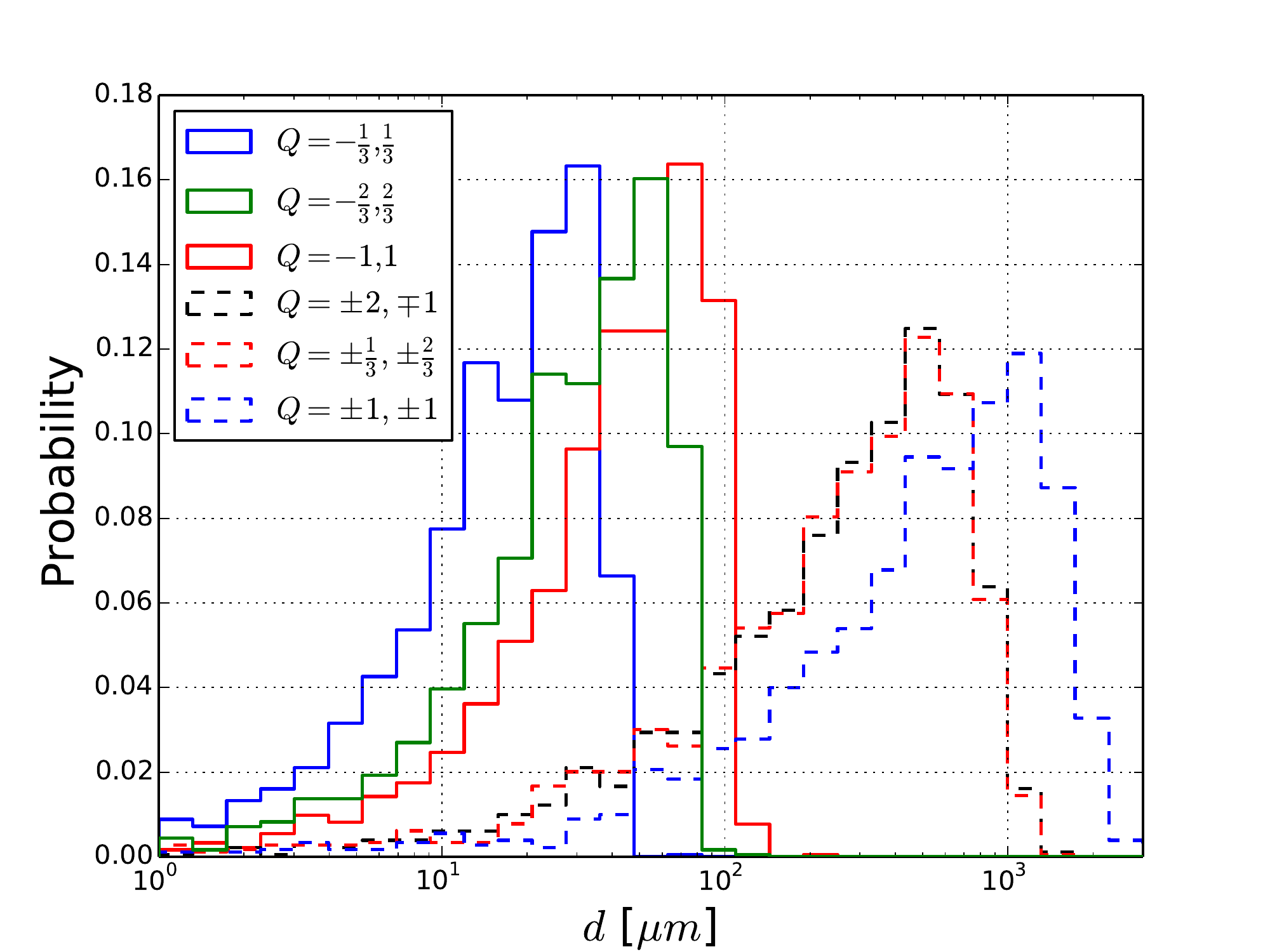}
\end{center}
\caption{Left: quirk pair plane thickness (calculated from numerical simulation instead of using Eq.~\ref{eq::dsim}) with varying quirk mass (${m}$), transverse momentum of quirk pair ($p_T$) and confinement scale ($\Lambda$). Right: quirk pair plane thickness for different quirk charges, where we choose ${m}=100$ GeV, $p_T>100$ GeV, {$\Lambda=\Lambda_0$}.   \label{fig::dnc} }
\end{figure}

In previous discussions, we have {chosen} the electric charge of quirks to be $\pm 1$ and the quirk-pair system is electric neutral.  However, our conclusion can be naturally applied to the quirks with different charges as long as the quirk pair is keeping neutral. Because it is the Lorentz force {rendering} the quirk traveling outside the plane, the plane thickness is linearly proportional to {each of the quirk charges.} This feature is clearly shown by the solid lines in the right panel of Fig.~\ref{fig::dnc}, where the thickness is calculated precisely from numerical simulation on the samples with different charges {while keeping the initial momentum and $\Lambda$ the same.} 

The quirk pair plane thickness will be dramatically increased for {non-neutral charge quirk-pair system}, since the trajectory of the quirk-pair system will be bended by the Lorentz force. {In this case} our discussion for the quirk pair plane thickness can only be used to roughly estimate the thickness increasement in one period of quirk's motion. As a result, the total quirk pair plane thickness will be increased {more intensely} by the number of periods inside the tracker, comparing to the thickness of electric neutral quirk-pair system. 
For the process with ${m}=100$ GeV, $p_T>100$ GeV, {$\Lambda=\Lambda_0$}, the number of periods of quirk motion inside tracker is around $20$, so the quirk pair with charges $\pm 1/3$ and $\pm 2/3$ should have plane thickness $\sim20$ times larger than quirk pair with charges $\pm 1/3(2/3)$ and $\mp 1/3(2/3)$, as demonstrated in the right panel of Fig.~\ref{fig::dnc}. 

\section{Crossing the same tracking layer more than once} \label{sec:morehit}
{The coplanar search proposed in Ref.~\cite{Knapen:2017kly} is designed to be 
based on the assumption that the quirk pair {induces} two hits on each barrel tracking layer. 
However, the quirk pair trajectories are highly dependent on the quirk mass, confinement scale 
as well as initial momentum. In many cases, the quirk pair may induce greater number of hits 
on tracking layers. Those intensive coplanar hits on single layer may serve as a useful handle to 
further suppress the backgrounds in the quirk search. 
We consider models with five different sets of parameters ($m, ~\Lambda,~p_T$). For each case, 
10K events are generated (assuming QCD production of quirk pair) to characterize the initial momentum distribution. 
Given an event in the CMS detector, different tracking layers can collect different numbers of hits. 
Among them, the largest one is recorded as $n^{\max}_{\text{hit}}$.
Tab.~\ref{tab:frachits} shows the fraction of events with a certain $n^{\max}_{\text{hit}}$ and in any event there is no minimum requirement on the number of hits in each layer.
Note that we use the notation F$_{n^{\max}_{\text{hit}}<2}=\text{F}_{n^{\max}_{\text{hit}}=0}+\text{F}_{n^{\max}_{\text{hit}}=1}$. In event with $n^{\max}_{\text{hit}}<2$, the quirk pair can only induce at most one hit on each tracking layer. 
Similarly,  F$_{n^{\max}_{\text{hit}}>4}=\text{F}_{n^{\max}_{\text{hit}}=5}+\text{F}_{n^{\max}_{\text{hit}}=6}+...+\text{F}_{n^{\max}_{\text{hit}}=\infty}$. 
We can see that there is {a} large fraction of events (typically around $10\%$) that will induce more than two hits in at least one tracking layer. This intensive hit fraction (F$_{n^{\max}_{\text{hit}}>2}$) is considerable for $\Lambda \gtrsim \mathcal{O}(1)$ keV. As for a given $\Lambda$, the fraction F$_{n^{\max}_{\text{hit}}>2}$ increases with increasing ${m}/p_T$ because of the increased quirk oscillation amplitude. 
}

\begin{table}[htb]
\begin{tabular}{|c|c|c||c|c|c|c|c|} \hline
${m}$ [GeV] & $p_T$ [GeV]&  $\Lambda$ [eV] & F$_{n^{\max}_{\text{hit}}<2}$ & F$_{n^{\max}_{\text{hit}}=2}$  & F$_{n^{\max}_{\text{hit}}=3}$  & F$_{n^{\max}_{\text{hit}}=4}$  & F$_{n^{\max}_{\text{hit}}>4}$ \\\hline
500 & 100 & 632.456 & 0.124 & 0.501 & 0.066 & 0.187 & 0.121 \\ \hline
100 & 100 & 632.456 & 0.114 & 0.744 & 0.021 & 0.102 & 0.019 \\ \hline
100 & 300 & 632.456 & 0.0076 & 0.933 & 0.0087 & 0.047 & 0.0034 \\\hline
100 & 100 & 200 & 0.123 & 0.780 & 0.032 & 0.055 & 0.009 \\ \hline
100 & 100 & 2000 & 0.113 & 0.586 & 0.009 & 0.192 & 0.099 \\ \hline
\end{tabular}
\caption{The fraction of quirk pair events that leave at most n hits on each tracking layer of the CMS detector is denoted by F$_{n^{\max}_{\text{hit}}=n}$. The different processes are characterized by fixing quirk mass ${m}$, confinement scale $\Lambda$ and least transverse momentum of quirk-pair system ($p_T$).  \label{tab:frachits}}
\end{table}

To be specific, {we use two parameters to characterize the shape of quirk pair trajectories. One is $L$ given in Eq.~\ref{eq::LL} and the other is defined as
\begin{align}
D= 2\frac{m}{\Lambda^2} \frac{\rho\beta_o}{\sqrt{1-\beta_o^2}} ~.
\end{align}
$L$ corresponds to the width of the belt which the tracks are traveling inside, and $D$ is the distance between two consecutive crossing points of two trajectories.} If we only consider the hits on barrel tracking layer, we can project the quirk trajectories onto the transverse plane. The $L$ and $D$ will be projected into $L^\prime=L \cos\phi$ and $D^\prime=D\sin\theta$. {In particular, }we demonstrate in the Appendix~\ref{app:hits} that the quirk pair can induce much more than 2 hits on a single tracking layer with radius $R$, if $R$ lies between $D'$ and $L'$, or $R$ is much {larger} than $L'$.

\section{Effects of ionization energy loss in mono-jet search} \label{sec:mono}
As pointed out in Ref.~\cite{Farina:2017cts} the non-helical trajectory of quirk will not be reconstructed in conventional searches at the LHC. So it will simply {represent} as missing transverse energy (MET) in event analyses. The mono-jet search at the LHC can {be used to constrain} the signal of quirk production with recoiling against a hard ISR jet.  

In fact, as studied in Ref.~\cite{Li:2019wce}, the quirk pair is not fully invisible. Since the quirk usually carries electric charge and travels with speed much smaller than the speed of light due to its heavy mass, it can deposit a certain amount of its energy inside the electromagnetic calorimeter (ECal) and  hadronic calorimeter (HCal). In the following, we will consider the effects of quirk ionization energy loss inside ECal and HCal on the selection efficiency of mono-jet search. 

\begin{figure}[thb]
\begin{center}
\includegraphics[width=0.45\textwidth]{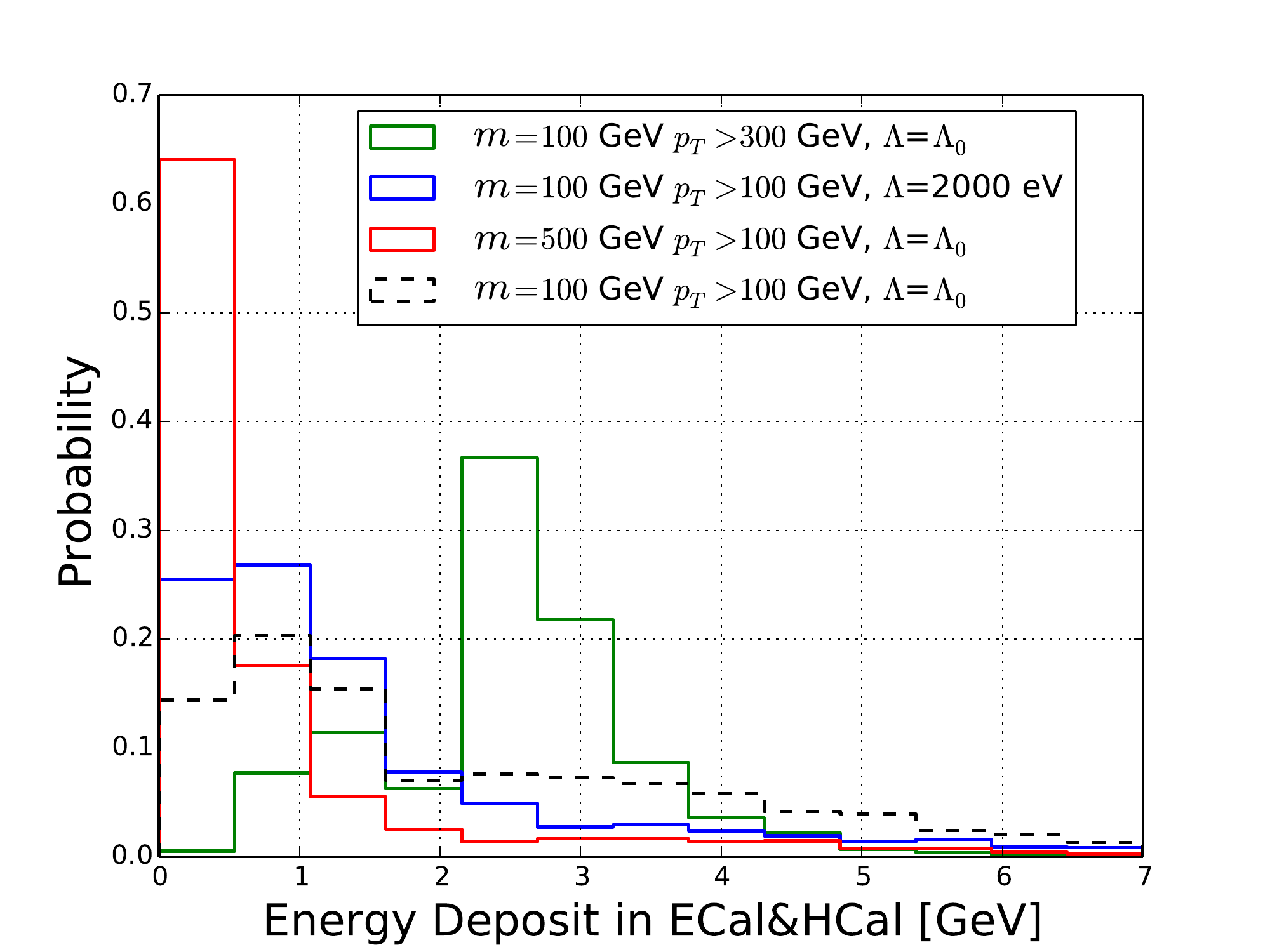}
\includegraphics[width=0.45\textwidth]{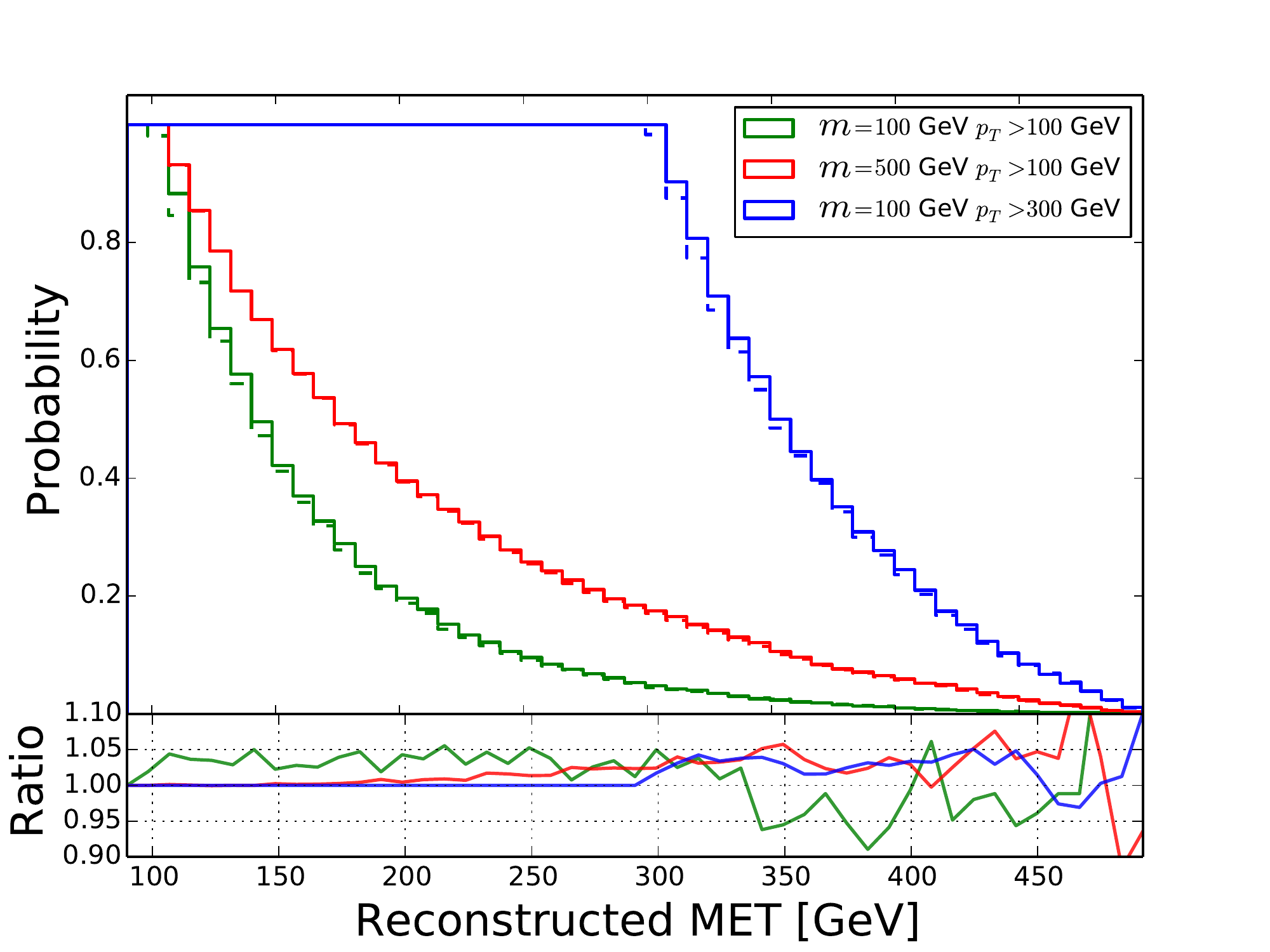}
\end{center}
\caption{\label{fig:monoj} Left: energy deposit of quirk pair in ECal and HCal within time of 25 ns. Right: the cumulative curve for the distributions of reconstructed MET. Note that the tails are dominated by the statistical fluctuation. The dashed lines have taken into account the quirk energy deposit.  }
\end{figure}

After numerical simulation as introduced in Ref.~\cite{Li:2019wce}, we can obtain the energy deposition in calorimeters of the CMS detector for different quirk production processes.  In principle, slower quirk tends to deposit more energy inside calorimeters, since the ionization energy loss is proportional to $\sim v^{-2}$(for velocity $v \gtrsim 0.1$). 
However, in our simulation, we only consider the energy deposition within 25 ns, which is the time interval of bunch crossing at the LHC. The slowly moving quirk-pair can not pass through the calorimeters in time. It leads to the low energy deposit for the quirk production process with ${m}=500$ GeV, $p_T>100$ GeV, 
and {$\Lambda=\Lambda_0$}, as shown in the left panel of Fig.~\ref{fig:monoj}. While the process with ${m}=100$ GeV, $p_T>100$ GeV, and {$\Lambda=\Lambda_0$} deposits largest energy because of its longest travel distance in the calorimeters. Note that larger $\Lambda$ {leads} to more accelerated quirk thus relatively smaller ionization energy loss. 

In experimental analyses, the MET of an event is reconstructed by using all energy deposits in the calorimeters. It means the quirk energy deposit will also be taken into account. As a result, the MET is overestimated if one {ignores} the quirk energy deposit, as done in Ref.~\cite{Farina:2017cts}. 
In the right panel of Fig.~\ref{fig:monoj}, we show how much the energy deposit of quirk will change the cut efficiency on MET, where the cumulative curves for the reconstructed MET of three different processes are shown. The dashed lines have taken into account the quirk energy deposits. In the lower subplot, the ratios between the MET without and with quirk energy deposits are given. We can see that cut efficiencies of MET is typically overestimated by a factor of 1.05, if the energy deposits of quirk are not included. {This turns out to be a small effect in practical analyses.}

\section{Variation of the $\hat{s}$ direction}  \label{sec:fixs}

In solving quirk EoM, one usually assumes the straight-string approximation~\cite{Kang:2008ea}, {\it i.e.},
 the infra-color string is straight at a given time in the CoM frame. 
In order to ensure the simultaneity in the CoM frame, the space-time position in the lab frame for two quirks ($t_{1,2},\vec{r}_{1,2}$) should satisfy
\begin{equation}\label{eq::synchrone}
t_1-t_2=\vec{\beta}\cdot(\vec{r}_1-\vec{r}_2).
\end{equation} 
It requires {that the time increasing step $\epsilon_{1,2}$ in numerical simulation {satisfies}}
\begin{align}\label{eq::length}
\epsilon_{1}[1- & \vec{v}_{1} \cdot \vec{\beta}-\frac{\vec{r}_{1}-\vec{r}_{2}}{E_{1}+E_{2}}\cdot(\vec{F}_{1}-\vec{v}_{1} \cdot \vec{F}_{1} \vec{\beta})]= 
\epsilon_{2}[1-\vec{v}_{2} \cdot \vec{\beta}-\frac{\vec{r}_{2}-\vec{r}_{1}}{E_{1}+E_{2}}\cdot(\vec{F}_{2}-\vec{v}_{2} \cdot \vec{F}_{2} \vec{\beta})]~,
\end{align}
where $\vec{F}_{i}=\vec{F}_{si}+\vec{F}_{exti}$ includes the infracolor force and external forces. 
{Then, at any time $t'_{1,2}$, the $\hat{s}_{1}$ and $\hat{s}_{2}$ used in Eq.~\ref{eq::fs} for two quirks in the lab frame are the unit vectors of  
\begin{align}
\vec{r}_{s1}&=(\vec{r}'_1-\vec{r}'_2)-(t'_1-t'_2)\vec{v}_1~,\label{eq::s1}\\
\vec{r}_{s2}&=(\vec{r}'_2-\vec{r}'_1)-(t'_2-t'_1)\vec{v}_2~,\label{eq::s2}
\end{align}
respectively. }

\begin{figure}[thb]
\begin{center}
\includegraphics[width=0.45\textwidth]{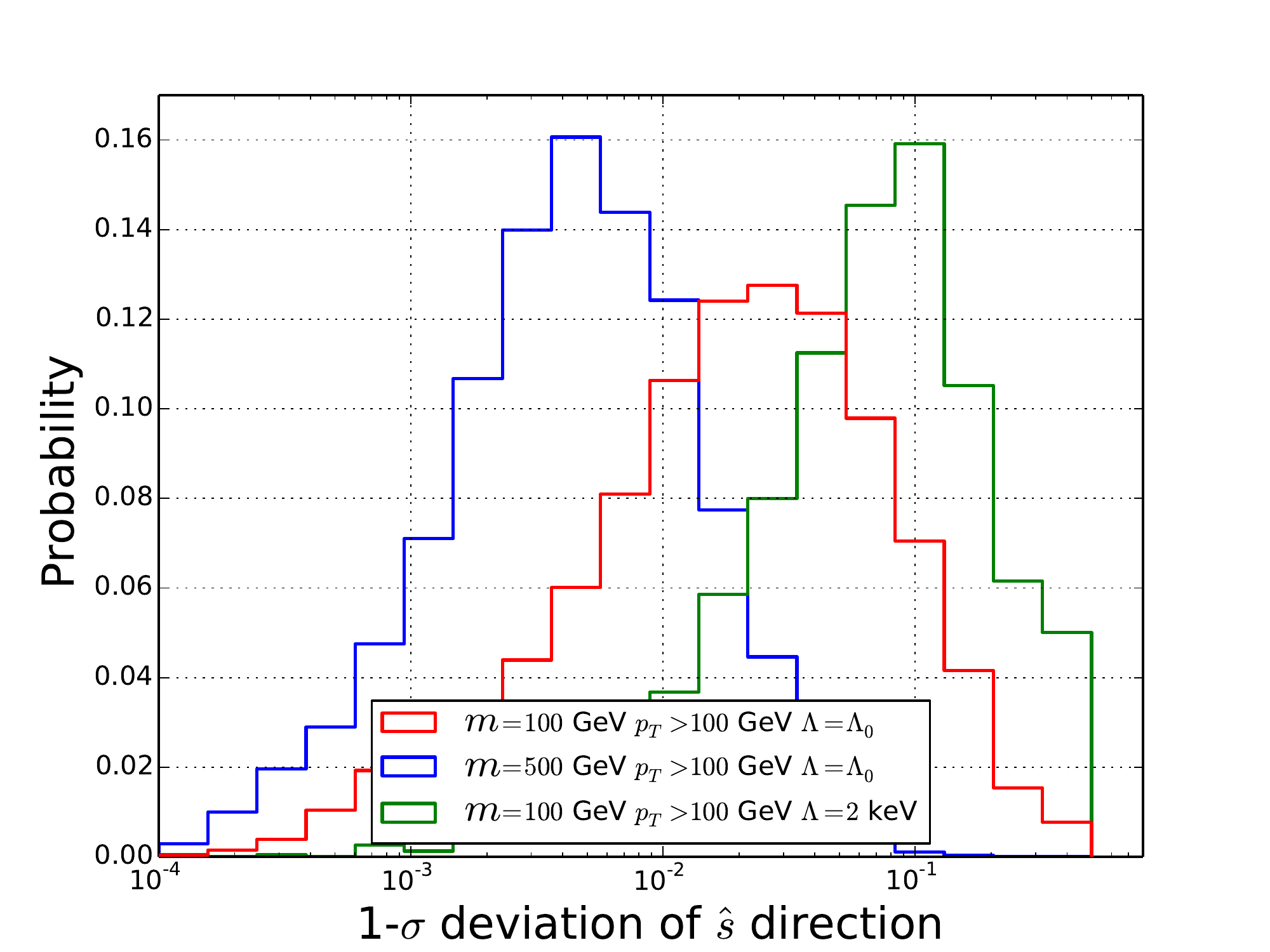}
\includegraphics[width=0.45\textwidth]{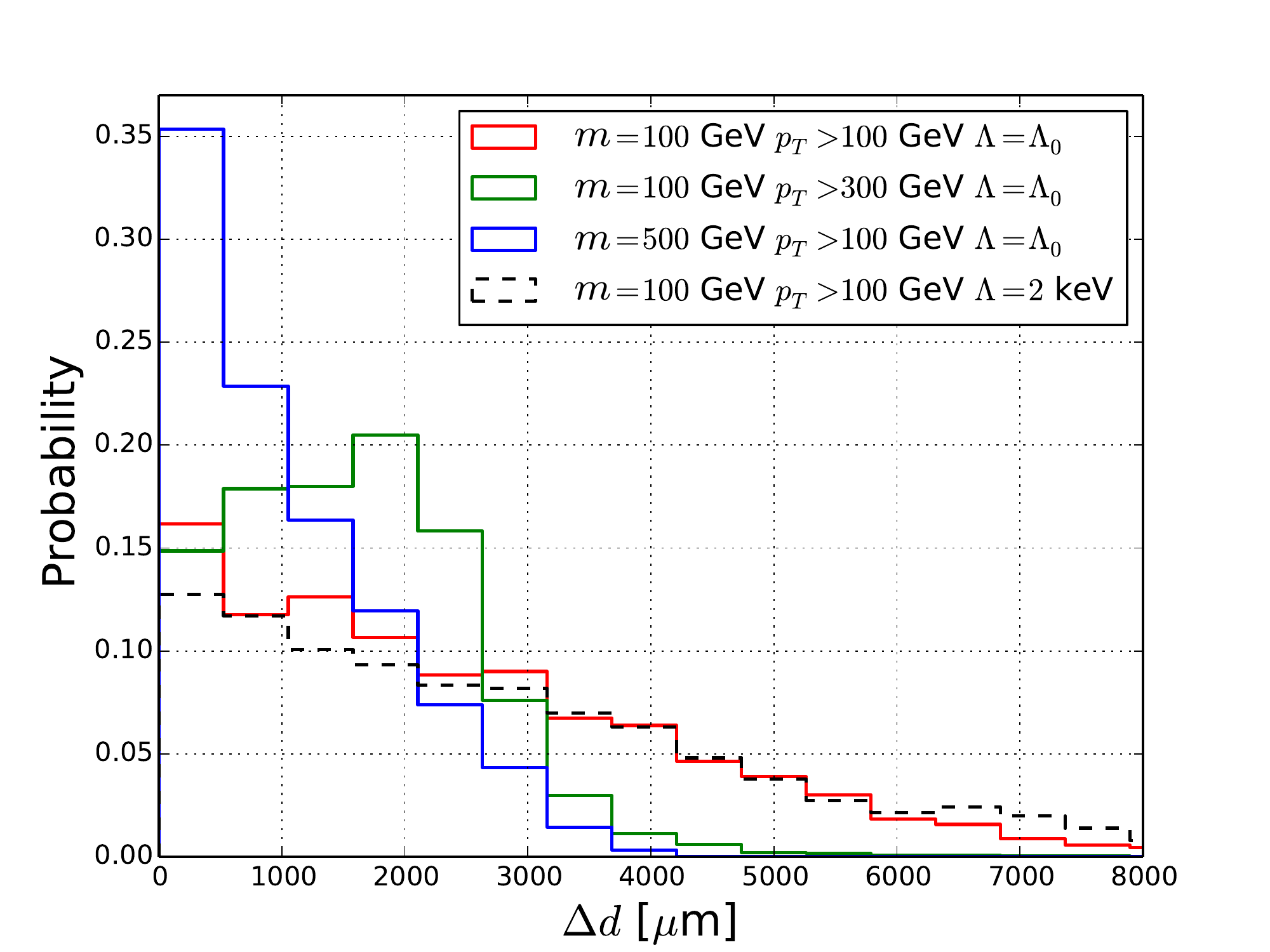}
\end{center}
\caption{\label{fig:vS} Left: the distributions of standard deviations of the $s$ angles for different quirk production processes. Right: the difference of quirk pair plane thickness between taking fixed $\hat{s}$ and true $\hat{s}$ respectively in solving quirk EoM. } 
\end{figure}

It is clear that the directions of $\hat{s}_{1,2}$ are varying at each time step of numerical solution. For each event, we can calculate the standard deviation ($\sigma(\hat{s})$) for {the angle between $\hat{s}$ at all time steps and $\pm \hat{e}_x$ (the initial $\hat{s}$) to characterize the varying range of $\hat{s}$.}  The distributions of $\sigma(\hat{s})$ for all events of several processes are shown in the left panel of Fig.~\ref{fig:vS}. The deviation of $\hat{s}$ increases as $\Lambda$ is increased, which can be around $\mathcal{O}(0.1)$ [rad] for $\Lambda \sim$ keV. Moreover, since the deviation is mainly induced by the Lorentz force, larger $p_T/{m}$ {renders} more significant deviation. 

Finally, we give a brief discussion on how a few existing quirk searches will change if one simply assumes the $\hat{s}$ is fixed (which seems taken in Ref.~\cite{Evans:2018jmd}). In the right panel of Fig.~\ref{fig:vS}, we plot the changes in the quirk pair plane thickness when the fixed $\hat{s}_{1,2}$ are used instead of 
Eqs.~\ref{eq::s1} and \ref{eq::s2}. The quirk pair plane thickness is changed dramatically (Note that a true quirk pair thickness is around $\mathcal{O}(100)$ $\mu m$ for our parameter choice). The influence is sensitive to the $p_T/m$ and {mildly depends} on the $\Lambda$. So choosing the correct $\hat{s}$ direction is critical in coplanar quirk search. 
On the other hand, we find that fixing $\hat{s}$ can only lead to at most 2-3\% changes in the cut efficiencies of mono-jet search. 
Moreover, each {quirk} trajectory can still be reconstructed approximately as helix with $\chi^2 <5$ when the confinement scale $\Lambda \lesssim \mathcal{O}(10)$ eV~\cite{Farina:2017cts}.  In this case, we find that fixing $\hat{s}$ can only change the $\chi^2$ by less than 1\% for ${m}=100$ GeV, $p_T>100$ GeV, and $\Lambda=10$ eV.

\section{Conclusion} \label{sec:conclude}

We solved the quirk equations of motion analytically in the limit of $|\vec{F}_{\text{ext}}|\ll\Lambda^2$, 
such that the external force can be treated as a correction to the infracolor force. 
According to the analytical solutions, the quirk pair oscillation amplitude can be expressed in a precise way in both CoM frame and laboratory frame. 
Meanwhile, the number of periods for quirk traveling inside tracker can be calculated immediately by using kinematic variables and model parameters, without conducting time consuming numerical simulation. 

The coplanar search proposed in Ref.~\cite{Knapen:2017kly} is one of the most efficient method for searching the quirk signal at colliders, when the confinement scale $\Lambda \in [\mathcal{O}(100)~\text{eV}, \mathcal{O}(10)~\text{keV}]$. We provided an approximate expression for the thickness of quirk pair plane in terms of kinematic variables and model parameters. 
Comparing with the precise numerical simulation results, we found the analytically expression can be valid up-to one order of magnitude if the number of quirk periods is between 1-100. 
This expression is especially useful to predict a model or parameter space which has large thickness of quirk pair plane, so that the coplanar search become less efficient. 
Also, we studied the electric charge dependence of the quirk pair plane thickness, and found that the plane thickness is linearly proportional to the quirk charges if the quirk-pair system is electric neutral, while the plane thickness will be increased by a factor of 
quirk period number if the quirk-pair system carries electric charge. 
The coplanar search becomes less efficient if the quirk crosses at least one of the tracking layers more than once.  The probability of this multi-crossing increases with increasing $\Lambda$ and ${m}/p_T$, which is typically $\sim \mathcal{O}(0.1)$ for the parameters of interest in this work. 

The effect of ionization energy loss inside the detector (including tracker, electromagnetic calorimeter, hadronic calorimeter and so on) is usually ignored in quirk signal analysis. We showed that this effect will lead to an overestimated MET cut efficiency by $\sim$ 5\%.
Moreover, the variation of the infracolor string direction $\hat{s}$ is typically small (much smaller than 0.1 [rad], depending on the kinematic variables and the $\Lambda$) for relatively small $\Lambda$. So it may be assumed that the direction is fixed in some analyses for simplification. We found that the correct direction is critical in coplanar quirk search. However, the mono-jet search and the heavy stable charged particle search are quite insensitive to the true $\hat{s}$.

\appendix
\section{Derivation for $z^{+}(g)$ and $z^{-}(g)$} \label{sec:zi}
\subsubsection{$z^{+}(g)$}

The acceleration related to $z^{+}(g)$ is
\begin{align}
\frac{d^2 z^{+}(g)}{d t_c^2}\hat{e}_z=\frac{q \vec{v}_{c1}(g)\times(B_c\hat{e}_{yc})}{m}{\sqrt{1-\vec{v}^2_{c1}(g)}}~,
\end{align}   
leading to
\begin{align}
z^{+}(g)&=\frac{1+(-1)^{[g]}}{2}z_o^{+}(\bar{g})+\frac{1-(-1)^{[g]}}{2}z_o^{+}(1-\bar{g}), \label{ap::zplus}\\
z_o^{+}(g)&=\frac{m q B_c}{\Lambda^4}\left(2\rho g- \tan^{-1}[\rho]+ \tan^{-1}[\rho(1-2g)]-\rho(1-2g)\ln[\sqrt{\frac{1+\rho^2}{1+\rho^2(1-2 g)^2}}] \right),\label{ap::zoplus}
\end{align}
with all relevant variables defined as in the main text. 

\subsubsection{$z^{-}(g)$}

According to Eqs.~\ref{ap::zplus} and \ref{ap::zoplus}
\begin{align}
z^{+}(2n)=0~,~~\frac{d z^{+}(2n)}{dt_c}=0~~~~~~~~ n=0,1,2\dots~,
\end{align}
which imply the total momentum of the quirk pair is invariant at $g=2n$. {Moreover, from Eq.~\ref{ap::rc}, we know the average torque on the system of two quirks is zero when $g$ is increased from $2n$ to $2n+2$ such that the total angular momentum of the quirk pair is also invariant at $g=2n$.} Then there must be 
\begin{align}
z_{1}(2n)=-z_{2}(2n)=z^{-}(2n)=0~~~~~~~~~n=0,1,2\dots~.
\end{align}
This means that two quirks meet each other and have opposite velocity at $g=2n$ {in the CoM frame}.
We hence conclude that kinematics of quirk system at $g+2$ can be obtained by rotating the system at $g$ with an angle $-\eta$ around $\hat{e}_{yc}$, leading to
\begin{align}
z^{-}(2n+\Delta)&=z^{-}(\Delta)\cos(n\eta)+(-1)^{[\Delta]} r(\bar{\Delta})\sin(n\eta)\nonumber\\
&\approx z^{-}(\Delta)+(-1)^{[\Delta]} r(\bar{\Delta}) n\eta~,\label{ap::zminus}
\end{align}
where $\Delta\in[0,2]$ and $\sin\eta=\frac{\sqrt{1+\rho^2}}{\rho}\frac{d z^{-}(g)}{dt_c}\Big{\lvert}_{g=2}\approx \eta\ll1$.
From Eq.~\ref{ap::vc}, 
\begin{align}
\vec{v}_{c1}(2n+\Delta)=\vec{v}_{c1}(2n-\Delta)~,
\end{align}
we have 
\begin{align}
z^{-}(2n+\Delta)-r(\Delta) n\eta= z^{-}(2n-\Delta)+r(\Delta) n\eta~,
\end{align}
and thus
\begin{align}\label{ap::zminus2}
z^{-}(\Delta)- z^{-}(2-\Delta)=r(\Delta)\eta~,
\end{align}
where $\Delta\in [0,1]$. We infer from  Eqs.~\ref{ap::zminus} and~\ref{ap::zminus2} that $z^{-}(g)$ can be formally written as
\begin{align}
z^{-}(g)&=\frac{1+(-1)^{[g]}}{2}\left( z_o^{-}(\bar{g})+\sin([\frac{g}{2}]\eta)r(\bar{g})\right) +\frac{1-(-1)^{[g]}}{2}\left( z_o^{-}(1-\bar{g})-\sin([\frac{g+1}{2}]\eta)r(\bar{g})\right)~. \label{ap::zominus}
\end{align} 
For $g\in [0,1]$, we have 
\begin{align}
\frac{d^2 z^{-}(g)}{d t_c^2}=\frac{d^2 z_o^{-}(g)}{d t_c^2}&\approx \frac{q E_z}{m}\sqrt{1-\vec{v}^2_{c1}(g)}~,
\end{align}
leading to
\begin{align}
z^{-}_o(g)&\approx z^{-}_{o\prime}(g)~,\\
z_{o\prime}^{-}(g)&=\frac{mq E_z}{\Lambda^4}\left(\sqrt{1+\rho^2}-\sqrt{1+\rho^2(1-2g)^2}-\rho(1-2g)\left( \sinh^{-1}[\rho]- \sinh^{-1}[\rho(1-2g)]\right)  \right)~. 
\end{align}
Results from the numerical calculations show that the difference between $z^{-}_o(g)$ and $z^{-}_{o\prime}(g)$ can not be ignored when $g\in [0.5,1]$, even though they are still of the same order of magnitude. Using
\begin{align}
z^{-}(0.5)&= w_{0.5}~,\\
z^{-}(1)&= w_{1}
\end{align}
from the numerical calculations, we can approximately construct $z_o^{-}(g)$ as  
\begin{align}
z_o^{-}(g)&=
\begin{cases}
\frac{w_{0.5}}{z_{o\prime}^-(0.5)}z_{o\prime}^-(g)~,~~~~~~~~~~~~~~~~~~~~~~~~~~~~~~~~~~~~~~~~~~g\in[0,0.5]\\
\left(\frac{w_{0.5}}{z_{o\prime}^-(0.5)} +(\frac{w_{1}}{z_{o\prime}^-(1)} -\frac{w_{0.5}}{z_{o\prime}^-(0.5)})(2g-1)\right)z_{o\prime}^-(g)~,~~~~g\in[0.5,1]
\end{cases}~.\label{eq::zo}
\end{align}

\subsubsection{Validation of the $z^{+}(g)$ and $z^{-}(g)$}

\begin{figure}[thb]
	\begin{center}
		\includegraphics[width=0.6\textwidth]{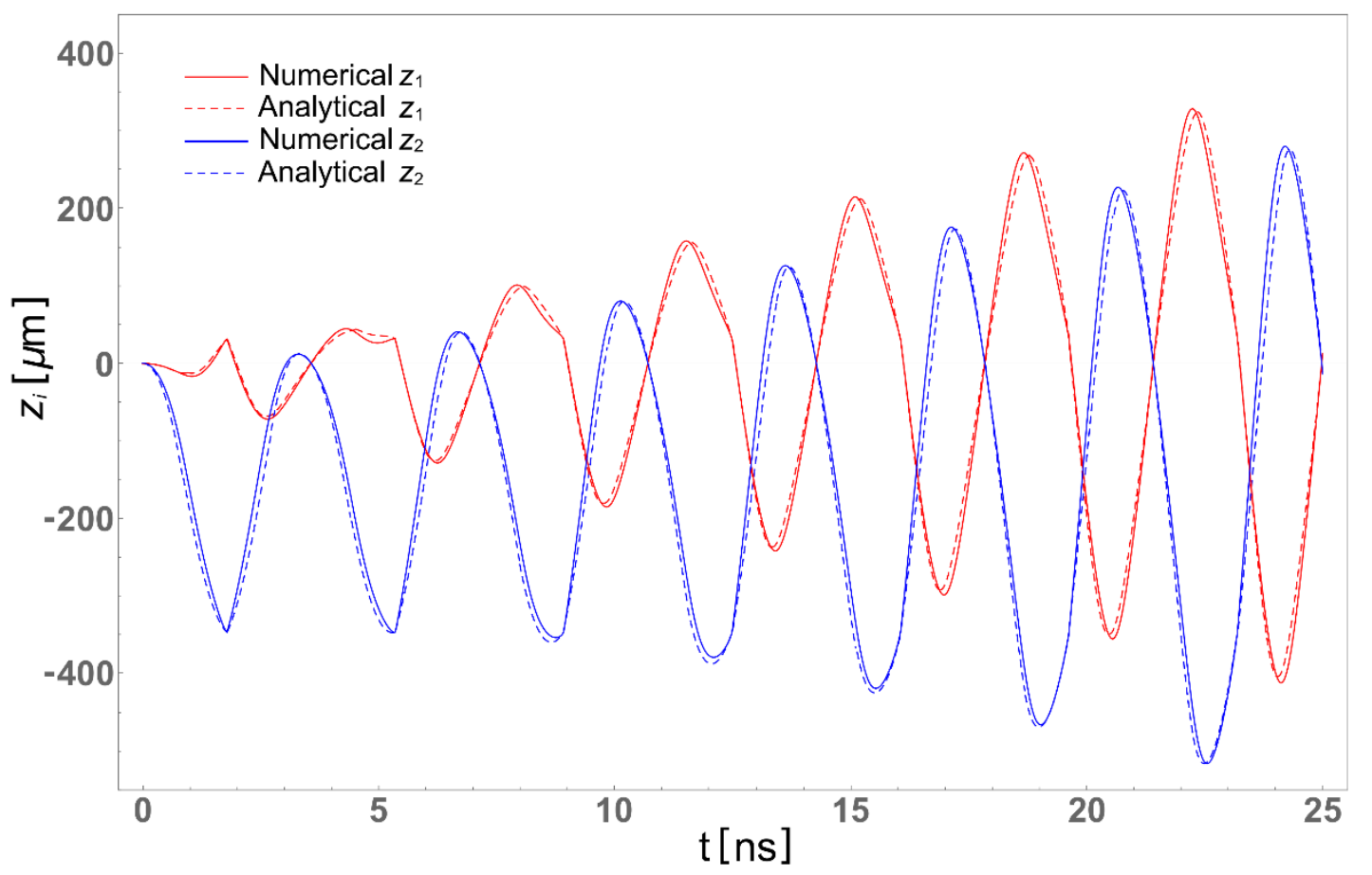}
	\end{center}
	\caption{$z_1$ and $z_2$ with respect to time variation obtained from analytical expressions as well as numerical simulation. For illustration, we choose $\vec{B}=(0,0,4)$ T in the lab frame, $\Lambda=500$ eV, quirks have charges of $\pm e$ and $m=100$ GeV, $\vec{P}_{o1}=(-279.6,250.7,4.8)$ GeV, and $\vec{P}_{o2}=(-36.5,102.5,303.6)$ GeV. }
	\label{fig::zmo}
\end{figure}

The above unknown parameters $w_{0.5}$, $w_1$  and {$\eta$} 
can be obtained from the numerical calculations. {Moreover, the magnitude order relation
\begin{align}\label{ap::magnitude}
r(0.5)|\eta| \sim |w_1| \sim |z^{-}_{o\prime}(1)|~, 
\end{align}
is also obtained by the numerical results.}
Following the definition, 
\begin{align}
z_1(g)& \equiv z^{+}(g)+z^{-}(g)~,\\
z_2(g)& \equiv z^{+}(g)-z^{-}(g)~
\end{align}
in the CoM frame, in Fig.~\ref{fig::zmo}, we make a comparison plot for $z_1(g)$ and $z_2(g)$ calculated 
from Eqs.~\ref{ap::zplus} and \ref{ap::zominus}, with those obtained by numerical simulations. It can be seen 
that our analytical expressions for $z_1(g)$ and $z_2(g)$ match the numerical results precisely. 

\section{Thickness of quirk pair plane}\label{sec:dc}
To simplify Eq.~\ref{eq:dc2}, we take the following approximations 
\begin{align}
&r(g)\approx 
\begin{cases}
2r(0.5)g~~~~~~~~~~~~~~g\in[0,0.5]\\
2r(0.5)(1-g)~~~~~~g\in[0.5,1]\\
\end{cases}~,\\
&z_o^{+}(g)\approx 
\begin{cases}
2z_o^{+}(1)g^2~~~~~~~~~~~~~~~~~~~~g\in[0,0.5]\\
z_o^{+}(1)\left( 1-2(1-g)^2\right)~~~~g\in[0.5,1]\\
\end{cases}~,\\
&z_o^{-}(g)\approx z_o^{-}(1)g^2~~~~~~g\in[0,1]~.
\end{align}
Then the conditions of
$\frac{\partial d_c^2}{\partial j}=
\frac{\partial d_c^2}{\partial c_z}=0$
{give}
\begin{align}
j(l_0)&=\frac{1}{l_0}\sum_{i=1}^{l_0}[\frac{i}{2}]~,\label{ap::j}\\
c_z&=\frac{z_o^{+}(1)}{2}~.\label{ap::cz}
\end{align}
Using Eq.~\ref{ap::j} and~\ref{ap::cz}, we obtain
\begin{align}
d_c^2&=C_1+C_2~, \label{eq:dc12}\\
C_1&=\frac{\eta^2r^2(0.5)}{3l_0}\sum_{i=1}^{l_0}\left( j(l_0)-[\frac{i}{2}]\right) ^2~,\\
C_2&=\frac{2z_o^{+ 2}(1)}{15}+\frac{z_{o}^{- 2}(1)}{5}~.
\end{align}

However, we know that the ideal quirk plane in the lab frame should also contain {the interaction point at which the quirk pair was produced}, which means we need to correct $C_2$ in $d_c^2$ (CoM frame thickness) to get $d^2$ (lab frame thickness).
For illustration, considering a rectangle which has length $L$ and  width $2\sqrt{C_2}$ ($2\sqrt{C_2}\ll L$), the diagonal line of the rectangle is the line that crosses one of the {vertexes} while having smallest distance square to the points on the edges. 
The corresponding average of the distance square is
$\frac{4}{3}C_2$ when $2\sqrt{C_2}\ll L$. So we multiply $C_2$ by $\frac{4}{3}$ in Eq.~\ref{eq:dc12} to get
\begin{align}
d^2&=\frac{\eta^2r^2(0.5)}{3l_0}\sum_{i=1}^{l_0}\left( j(l_0)-[\frac{i}{2}]\right) ^2+\frac{4}{3}\left( \frac{2{z_o^{+ }}^2(1)}{15}+\frac{{z_{o}^{-}}^2(1)}{5}\right)~. \label{eq::dd}
\end{align}

\section{Validation of the quirk pair plane thickness} \label{sec:dvalid}

\begin{figure}[thb]
\begin{center}
\includegraphics[width=0.3\textwidth]{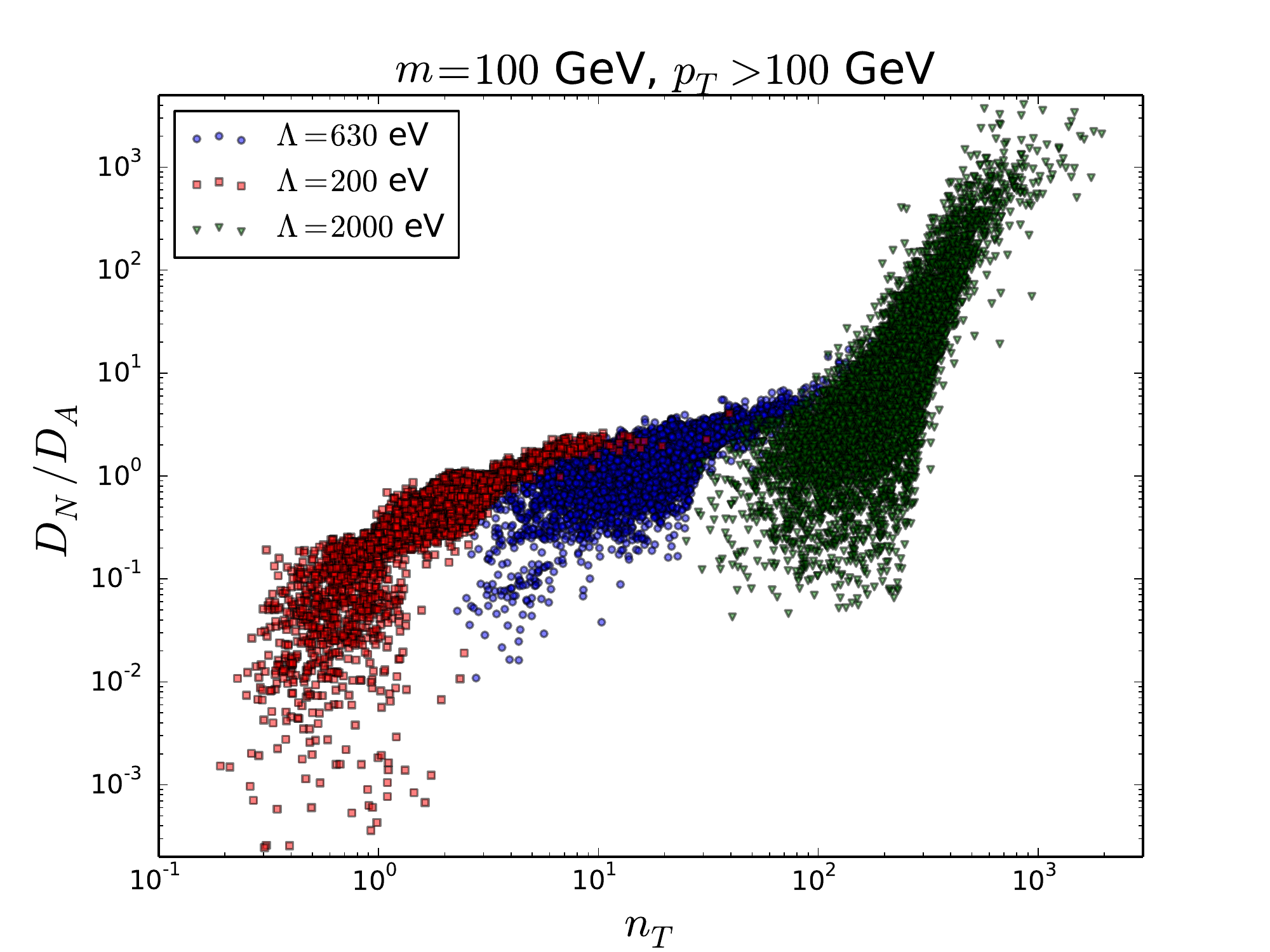}
\includegraphics[width=0.3\textwidth]{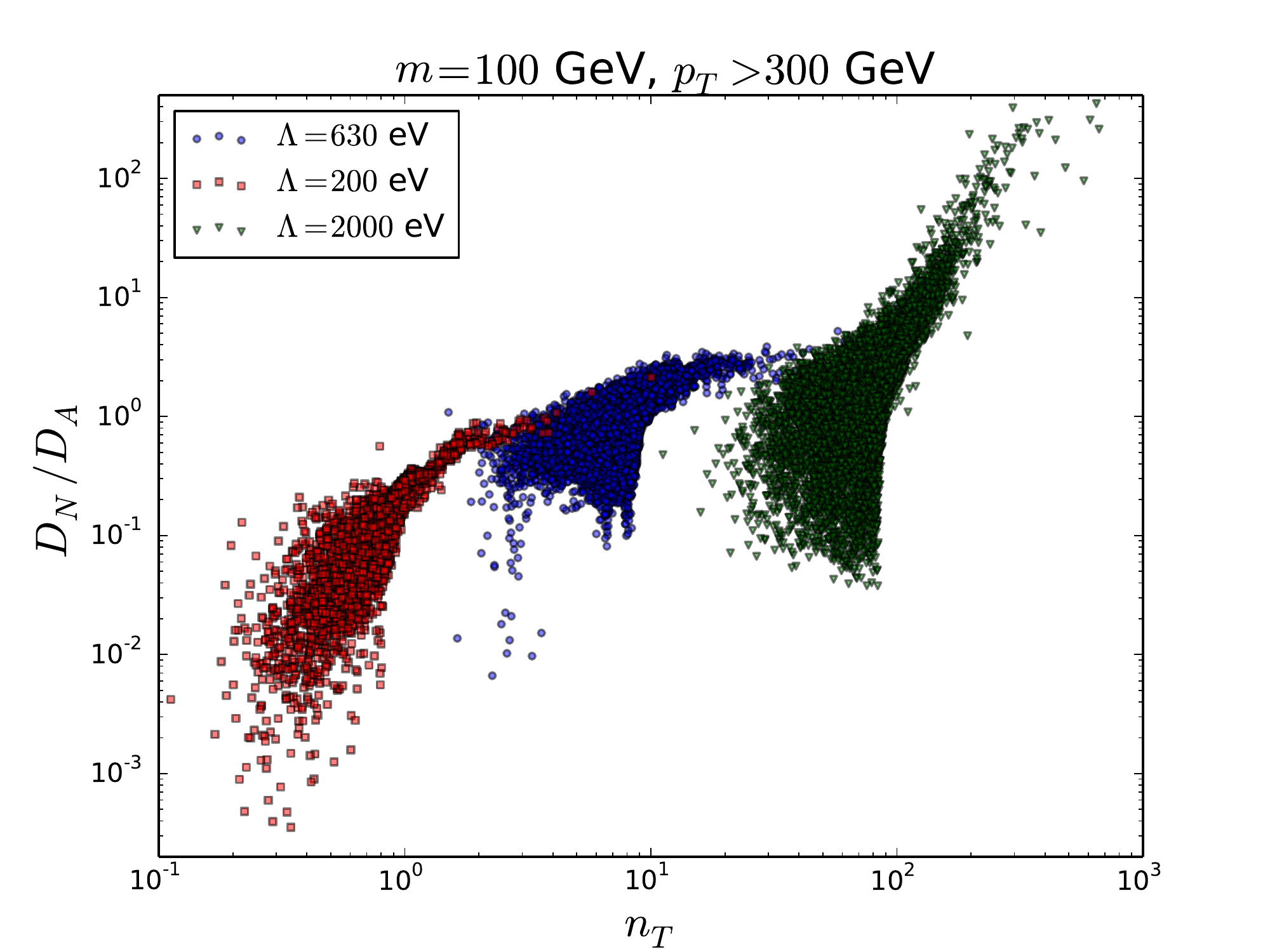}
\includegraphics[width=0.3\textwidth]{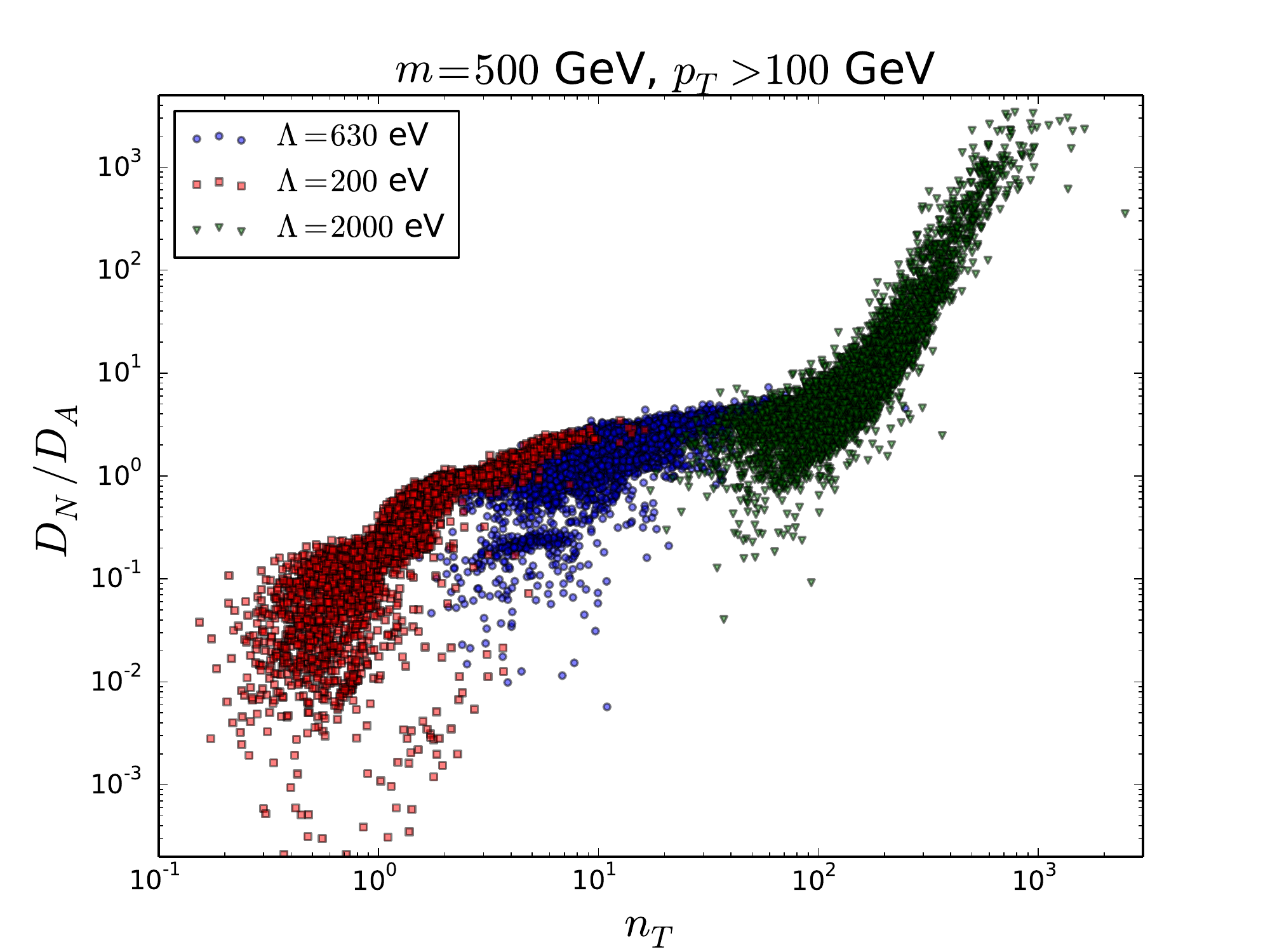}
\end{center}
\caption{\label{fig:dvalid} The ratio between the quirk pair plane thickness obtained from numerical simulation $(D_N)$ and that calculated from Eq.~\ref{eq::dsim} ($D_A$), with respect to the quirk trajectory periods ($n_T$) inside the tracker.  }
\end{figure}

On one hand, we can use Eq.~\ref{eq::dsim} to estimate the quirk pair plane thickness approximately with initial quirk kinematics. On the other hand, a more precise but time consuming way to obtain the thickness will be simulating the tracker configuration according to a specific detector and solving the quirk EoM numerically. Then, the quirk pair plane thickness square ($(D_N)^2$) corresponds to the smallest eigenvalue of the two-tensor~\cite{Knapen:2017kly}
\begin{align}
T(\vec{h}_a)_{ij} = \frac{1}{N-1} \sum_{a=1}^{N} \vec{h}^a_i \vec{h}^a_j~, 
\end{align}
where $\vec{h}_a$ is the position of $a$th hit in the tracker caused by the quirk pair. 
In Fig.~\ref{fig:dvalid}, we plot the ratio between quirk pair plane thickness obtained from numerical simulation and that calculated from Eq.~\ref{eq::dsim}. The analytical result matches the numerical result within an order of magnitude when the quirk trajectory period ($n_T$) inside the tracker is around $\mathcal{O}(1-100)$.
{The case with $n_T<1$ can not be {described} by Eq.~\ref{eq::dd}, since the derivations in Sec.~\ref{sec:thickness} 
use the oscillatory feature of the quirk motion which means that the number of periods needs to be larger than one. The quirk plane designed in Sec.~\ref{sec:thickness} does not give the smallest plane thickness when $n_T<1$ and thus the plane thickness given by Eq.~\ref{eq::dd} is larger than the numerical result. Strictly speaking, $E_z$ and $B_c$ respectively in Eqs.~\ref{ap::de} and \ref{ap::db} change with the number of periods due to the rotation of the quirk-pair system described by Eq.~\ref{ap::zominus}, but are thought to be invariant because of the small $\eta$. The effect of the system rotation can not be ignored  when $n_T>\mathcal{O}(100)$, so the plane thickness given by Eq.~\ref{eq::dd} is less than the numerical result. }

\section{More than two hits on single tracking layer} \label{app:hits}
Here, we provide a benchmark study to count the number of hits on each tracking layer to illustrate our discussions in Sec.~\ref{sec:morehit}. 
The benchmark point is chosen as ${m}=100$ GeV, $k_1$=12.22, $k_2$=12.86, $\alpha$=2.81, $\theta$=1.06 and $\phi$=0.72. The confinement scale  $\Lambda$ is varying from 400 eV to 3 keV. 
In Tab.~\ref{tab:nhits}, we show the number of hits produced by the quirk pair on each tracking layer (characterized by the radius $R$ where we have adopted the CMS detector configuration). On the left part of the table, we also give the corresponding values for $L'$ and $D'$ that are defined in Sec.~\ref{sec:morehit}. 
This benchmark study clearly shows that the quirk pair can induce more than 2 hits on the tracking layer with radius $R$ lying between $D'$ and $L'$. 

\begin{table}[htbp]
	\centering
	\begin{tabular}{|c|c|c||c|c|c|c|c|c|c|c|c|c|c|c|c|}
	\hline 
	$L^\prime$ [cm] & $D^\prime$ [cm] & \diagbox[width=0.12\textwidth]{$\Lambda$ [eV] }{R [cm]} & 4.4 & 7.3 & 10.2 & 25.5 & 33.9 & 41.85 & 49.8 & 60.8 & 69.2 & 78.0 & 86.8 & 96.5 & 108.0\\ \hline
	%$\Lambda$ & $L^\prime$ & $D^\prime$ & 4.4 & 7.3 & 10.2 & 25.5 & 33.9 & 41.85 & 49.8 & 60.8 & 69.2 & 78.0 & 86.8 & 96.5 & 108.0\\
	%eV & cm & cm & cm & cm & cm & cm & cm & cm & cm & cm & cm & cm & cm& cm & cm\\
	\hline 
	104.72 & 45.14 &400 & 2 & 2 & 2 
	& 2 & 2 & 4 & 6 
	& 6 & 6 & 6 & 8 & 8 & 5\\ 
	\hline 
	82.74 & 35.66 &450 & 2 & 2 & 2 
	& 2 & 4 & 6 & 6 
	& 6 & 8 & 10 & 8 & 4 & 5\\ 
	\hline 
	67.02 & 28.89 &500 & 2 & 2 & 2 
	& 2 & 6 & 6 & 6 
	& 10 & 8 & 4 & 8 & 4 & 3\\ 
	\hline 
	46.54 & 20.06 &600 & 2 & 2 & 2 
	& 6 & 6 & 10 & 8 
	& 6 & 4 & 2 & 4 & 2 & 4\\ 
	\hline 
	34.19 & 14.74 &700 & 2 & 2 & 2 
	& 6 & 10 & 6 & 4 
	& 4 & 2 & 4 & 2 & 2 & 3\\ 
	\hline 
	 26.18 & 11.28 &800 & 2 & 2 & 2 
	& 10 & 8 & 6 & 4 
	& 4 & 4 & 2 & 2 & 2 & 1\\ 
	\hline 
	 20.68 & 8.92 &900 & 2 & 2 & 6 
	& 6 & 4 & 2 & 2 
	& 2 & 2 & 2 & 2 & 2 & 3\\ 
	\hline 
	 16.75 & 7.22 &1000 & 2 & 6 & 6 
	& 4 & 2 & 2 & 2 
	& 2 & 2 & 2 & 4 & 2 & 3\\ 
	\hline
	 7.45 & 3.21 &1500 & 6 & 10 & 4 
	& 4 & 2 & 4 & 2 
	& 2 & 4 & 4 & 2 & 4 & 0\\ 
	\hline
	 4.19 & 1.81 &2000 & 8 & 4 & 2 
	& 2 & 2 & 2 & 4 
	& 4 & 2 & 2 & 2 & 0 & 0\\ 
	\hline
	 2.68 & 1.16 &2500 & 4 & 4 & 4 
	& 2 & 2 & 2 & 4 
	& 4 & 4 & 6 & 0 & 0 & 0\\ 
	\hline
	 1.86 & 0.80 &3000 & 4 & 2 & 2 
	& 4 & 4 & 4 & 6 
	& 0 & 0 & 0 & 0 & 0 & 0\\ 
	\hline
	\end{tabular}
	\caption{Number of hits in different cylindrical barrels of CMS tracker induced by the quirk pair within 25 ns. We also provide the corresponding $L'$ and $D'$ values in the left part of the table. }
	\label{tab:nhits}
\end{table}

\newpage
\section*{Acknowledgement}
This work was supported in part by the Fundamental Research Funds for the Central Universities, 
 by the NSFC under grant No. 11905149,  by
the Projects 11875062 and 11947302 supported by the 
National Natural Science Foundation of China, and by 
the Key Research Program of Frontier Science, CAS.

%%%%%%%%%%%%%%%%%%%%%
\phantomsection
\addcontentsline{toc}{section}{References}
\bibliographystyle{jhep}
\bibliography{quirk}

\end{document}